\documentclass[useAMS,usenatbib,breaklinks=true]{mnras}
\usepackage{graphicx}
\usepackage{times}
\usepackage{amssymb}
\usepackage{amsmath}
\newcommand{\mpcoh}{\,h^{-1}\,{\rm Mpc}}

\usepackage{hyperref}
\hypersetup{
    colorlinks=true,
    linkcolor=blue,
    citecolor=blue,
}

\def\ie{{\frenchspacing\it i.e.}}

\def\be{\begin{equation}}
\def\ee{\end{equation}}
\def\ba{\begin{eqnarray}}
\def\ea{\end{eqnarray}}

\def\rr{{\bf r}}

\def\k{{\bf k}}

\begin{document}

% --- title --- %
\title[DR14 eBOSS Quasar BAO Measurements] 
{The clustering of the SDSS-IV extended Baryon Oscillation Spectroscopic Survey DR14 quasar sample: First measurement of Baryon Acoustic Oscillations between redshift 0.8 and 2.2}

\author[M. Ata et al.]{\parbox{\textwidth}{\Large
Metin Ata$^{1}$,
Falk Baumgarten$^{1,2}$,
Julian Bautista$^{3}$,
Florian Beutler$^{4}$,
Dmitry Bizyaev$^{5,6}$,
Michael R. Blanton$^{7}$,
Jonathan A. Blazek$^{8}$,
Adam S. Bolton$^{3,9}$,
Jonathan Brinkmann$^{5}$,
Joel R. Brownstein$^{3}$,
Etienne Burtin$^{10}$,
Chia-Hsun Chuang$^{1}$,
Johan Comparat$^{11}$,
Kyle S. Dawson$^{3}$,
Axel de la Macorra$^{12}$,
Wei Du$^{13}$,
H\'elion du Mas des Bourboux$^{10}$,
Daniel J. Eisenstein$^{14}$,
H\'ector Gil-Mar\'in\thanks{Email: hector.gilmarin@lpnhe.in2p3.fr}$^{15,16}$,
Katie Grabowski$^{5}$,
Julien Guy$^{17}$,
Nick Hand$^{18}$,
Shirley Ho$^{19,20,21}$,
Timothy A. Hutchinson$^{3}$,
Mikhail M. Ivanov$^{22,23}$,
Francisco-Shu Kitaura$^{24,25}$,
Jean-Paul Kneib$^{8,26}$,
Pierre Laurent$^{10}$,
Jean-Marc Le Goff$^{10}$,
Joseph E. McEwen$^{27}$,
Eva-Maria Mueller$^{4}$,
Adam D. Myers$^{3}$,
Jeffrey A. Newman$^{28}$,
Nathalie Palanque-Delabrouille$^{10}$,
Kaike Pan$^{5}$,
Isabelle P\^aris$^{26}$,
Marcos Pellejero-Ibanez$^{24,25}$,
Will J. Percival$^{4}$,
Patrick Petitjean$^{29}$,
Francisco Prada$^{30,31,32}$,
Abhishek Prakash$^{28}$,
Sergio A. Rodr\'iguez-Torres$^{30,31,33}$,
Ashley J. Ross\thanks{Email: Ashley.Jacob.Ross@gmail.com}$^{27,4}$,
Graziano Rossi$^{34}$,
Rossana Ruggeri$^{4}$,
Ariel G. S\'anchez$^{11}$,
Siddharth Satpathy$^{20,35}$,
David J. Schlegel$^{19}$,
Donald P. Schneider$^{36,37}$,
Hee-Jong Seo$^{38}$,
An\v{z}e Slosar$^{39}$,
Alina Streblyanska$^{24,25}$,
Jeremy L. Tinker$^{7}$,
Rita Tojeiro$^{40}$,
Mariana Vargas Maga\~na$^{12}$,
M. Vivek$^{3}$,
Yuting Wang$^{13,4}$,
Christophe Y\`eche$^{10,19}$,
Liang Yu$^{41}$,
Pauline Zarrouk$^{10}$,
Cheng Zhao$^{41}$,
Gong-Bo Zhao\thanks{Email: gbzhao@nao.cas.cn}$^{13,4}$,
Fangzhou Zhu$^{42}$
 } \vspace*{4pt} \\ 
\scriptsize $^{1}$ Leibniz-Institut f\"{u}r Astrophysik Potsdam (AIP), An der Sternwarte 16, D-14482 Potsdam, Germany\vspace*{-2pt} \\ 
\scriptsize $^{2}$ Humboldt-Universit\"at zu Berlin, Institut f\"ur Physik, Newtonstrasse 15,D-12589, Berlin, Germany\vspace*{-2pt} \\ 
\scriptsize $^{3}$ Department Physics and Astronomy, University of Utah, 115 S 1400 E, Salt Lake City, UT 84112, USA\vspace*{-2pt} \\ 
\scriptsize $^{4}$ Institute of Cosmology \& Gravitation, Dennis Sciama Building, University of Portsmouth, Portsmouth, PO1 3FX, UK\vspace*{-2pt} \\ 
\scriptsize $^{5}$ Apache Point Observatory and New Mexico State University, P.O. Box 59, Sunspot, NM 88349, USA\vspace*{-2pt} \\ 
\scriptsize $^{6}$ Sternberg Astronomical Institute, Moscow State University, Universitetski pr. 13, 119992 Moscow, Russia\vspace*{-2pt} \\ 
\scriptsize $^{7}$ Center for Cosmology and Particle Physics, Department of Physics, New York University, New York, NY 10003, USA\vspace*{-2pt} \\ 
\scriptsize $^{8}$ Institute of Physics, Laboratory of Astrophysics, \'Ecole Polytechnique F\'ed\'erale de Lausanne (EPFL), Observatoire de Sauverny, 1290 Versoix, Switzerland\vspace*{-2pt} \\ 
\scriptsize $^{9}$ National Optical Astronomy Observatory, 950 N Cherry Ave, Tucson, AZ 85719, USA\vspace*{-2pt} \\ 
\scriptsize $^{10}$ IRFU,CEA, Universit\'e Paris-Saclay, F-91191 Gif-sur-Yvette, France\vspace*{-2pt} \\ 
\scriptsize $^{11}$ Max-Planck-Institut f\"ur Extraterrestrische Physik, Postfach 1312, Giessenbachstr., 85748 Garching bei M\"unchen, Germany\vspace*{-2pt} \\ 
\scriptsize $^{12}$ Instituto de F\'isica, Universidad Nacional Aut\'onoma de M\'exico, Apdo. Postal 20-364, M\'exico\vspace*{-2pt} \\ 
\scriptsize $^{13}$ National Astronomy Observatories, Chinese Academy of Science, Beijing, 100012, P.R. China\vspace*{-2pt} \\ 
\scriptsize $^{14}$ Harvard-Smithsonian Center for Astrophysics, 60 Garden St., MS \#20, Cambridge, MA 02138, USA\vspace*{-2pt} \\ 
\scriptsize $^{15}$ Sorbonne Universit\'es, Institut Lagrange de Paris (ILP), 98 bis Boulevard Arago, 75014 Paris, France\vspace*{-2pt} \\ 
\scriptsize $^{16}$ Laboratoire de Physique Nucl\'eaire et de Hautes Energies, Universit\'e Pierre et Marie Curie, 4 Place Jussieu, 75005 Paris, France\vspace*{-2pt} \\ 
\scriptsize $^{17}$ LPNHE, CNRS/IN2P3, Universit\'e Pierre et Marie Curie Paris 6, Universit\'e Denis Diderot Paris, 4 place Jussieu, 75252 Paris CEDEX, France\vspace*{-2pt} \\ 
\scriptsize $^{18}$ Department of Astronomy, University of California at Berkeley, Berkeley, CA 94720, USA\vspace*{-2pt} \\ 
\scriptsize $^{19}$ Lawrence Berkeley National Laboratory, 1 Cyclotron Road, Berkeley, CA 94720, USA\vspace*{-2pt} \\ 
\scriptsize $^{20}$ Department of Physics, Carnegie Mellon University, 5000 Forbes Avenue, Pittsburgh, PA 15213, USA\vspace*{-2pt} \\ 
\scriptsize $^{21}$ Berkeley Center for Cosmological Physics, LBL and Department of Physics, University of California, Berkeley, CA 94720, USA\vspace*{-2pt} \\ 
\scriptsize $^{22}$ Institute of Physics, LPPC, \'Ecole Polytechnique F\'ed\'erale de Lausanne (EPFL), CH-1015, Lausanne, Switzerland\vspace*{-2pt} \\ 
\scriptsize $^{23}$ Institute for Nuclear Research of the Russian Academy of Sciences, 60th October Anniversary Prospect, 7a, 117312 Moscow, Russia\vspace*{-2pt} \\ 
\scriptsize $^{24}$ Instituto de Astrof\'isica de Canarias (IAC), C/V\'ia L\'actea, s/n, E-38200, La Laguna, Tenerife, Spain\vspace*{-2pt} \\ 
\scriptsize $^{25}$ Dpto. Astrof\'isica, Universidad de La Laguna (ULL), E-38206 La Laguna, Tenerife, Spain\vspace*{-2pt} \\ 
\scriptsize $^{26}$ Aix-Marseille Universit\'e, CNRS, LAM (Laboratoire d'Astrophysique de Marseille), 38 rue F. Joliot-Curie 13388 Marseille Cedex 13, France\vspace*{-2pt} \\ 
\scriptsize $^{27}$ Center for Cosmology and Astro-Particle Physics, Ohio State University, Columbus, Ohio, USA\vspace*{-2pt} \\ 
\scriptsize $^{28}$ Department of Physics and Astronomy and the Pittsburgh Particle Physics, Astrophysics and Cosmology Center (PITT PACC), University of Pittsburgh, 3941 O'Hara Street, Pittsburgh, PA 15260, USA\vspace*{-2pt} \\ 
\scriptsize $^{29}$ Institut d'Astrophysique de Paris, Universit\'e Paris 6 et CNRS, 98bis Boulevard Arago, 75014 Paris, France\vspace*{-2pt} \\ 
\scriptsize $^{30}$ Instituto de F\'isica Te\'orica (UAM/CSIC), Universidad Aut\'onoma de Madrid, Cantoblanco, E-28049 Madrid, Spain\vspace*{-2pt} \\ 
\scriptsize $^{31}$ Campus of International Excellence UAM+CSIC, Cantoblanco, E-28049 Madrid, Spain\vspace*{-2pt} \\ 
\scriptsize $^{32}$ Instituto de Astrof\'isica de Andaluc\'ia (CSIC), E-18080 Granada, Spain\vspace*{-2pt} \\ 
\scriptsize $^{33}$ Departamento de F\'isica Te\'orica M8, Universidad Aut\'onoma de Madrid, E-28049 Cantoblanco, Madrid, Spain\vspace*{-2pt} \\ 
\scriptsize $^{34}$ Department of Physics and Astronomy, Sejong University, Seoul 143-747, Korea\vspace*{-2pt} \\ 
\scriptsize $^{35}$ The McWilliams Center for Cosmology, Carnegie Mellon University, 5000 Forbes Ave., Pittsburgh, PA 15213, USA\vspace*{-2pt} \\ 
\scriptsize $^{36}$ Department of Astronomy and Astrophysics, The Pennsylvania State University, University Park, PA 16802, USA\vspace*{-2pt} \\ 
\scriptsize $^{37}$ Institute for Gravitation and the Cosmos, The Pennsylvania State University, University Park, PA 16802, USA\vspace*{-2pt} \\ 
\scriptsize $^{38}$ Department of Physics and Astronomy, Ohio University, 251B Clippinger Labs, Athens, OH 45701, USA\vspace*{-2pt} \\ 
\scriptsize $^{39}$ Brookhaven National Laboratory, Bldg 510, Upton, New York 11973, USA\vspace*{-2pt} \\ 
\scriptsize $^{40}$ School of Physics and Astronomy, University of St Andrews, St Andrews, KY16 9SS, UK\vspace*{-2pt} \\ 
\scriptsize $^{41}$ Tsinghua Center for Astrophysics and Department of Physics, Tsinghua University, Beijing 100084, China\vspace*{-2pt} \\ 
\scriptsize $^{42}$ Department of Physics, Yale University, 260 Whitney Ave, New Haven, CT 06520, USA\vspace*{-2pt} \\ 
}

\date{} 

\pagerange{\pageref{firstpage}--\pageref{lastpage}} \pubyear{2017}
\maketitle
\label{firstpage}

\begin{abstract}
We present measurements of the Baryon Acoustic Oscillation (BAO) scale in redshift-space using the clustering of quasars. We consider a sample of 147,000 quasars from the extended Baryon Oscillation Spectroscopic Survey (eBOSS) distributed over 2044 square degrees with redshifts $0.8 < z < 2.2$ and measure their spherically-averaged clustering in both configuration and Fourier space.  Our observational dataset and the 1400 simulated realizations of the dataset 
allow us to detect a preference for BAO that is greater than 2.8$\sigma$. We determine the spherically averaged BAO distance to $z = 1.52$ to 3.8 per cent precision: $D_V(z=1.52)=3843\pm147 \left(r_{\rm d}/r_{\rm d, fid}\right)\ $Mpc. This is the first time the location of the BAO feature has been measured between redshifts 1 and 2. Our result is fully consistent with the prediction obtained by extrapolating the Planck flat $\Lambda$CDM best-fit cosmology. All of our results are consistent with basic large-scale structure (LSS) theory, confirming quasars to be a reliable tracer of LSS, and provide a starting point for numerous cosmological tests to be performed with eBOSS quasar samples. We combine our result with previous, independent, BAO distance measurements to construct an updated BAO distance-ladder. Using these BAO data alone and marginalizing over the length of the standard ruler, we find $\Omega_{\Lambda} > 0$ at 6.6$\sigma$ significance when testing a $\Lambda$CDM model with free curvature.
\end{abstract}


%section
\section{Introduction}

Using Baryon Acoustic Oscillations (BAOs) to measure the expansion of the Universe is now a mature field, with the BAO signal having been detected and measured to ever greater precision using data from a number of large galaxy surveys including: the Sloan Digital Sky Survey (SDSS) I and II (e.g., \citealt{Eisenstein05,Percival10,Ross15}), the 2-degree Field Galaxy Redshift Survey (2dFGRS) \citep{Percival01,2dF}, WiggleZ \citep{Blake11}, and the 6-degree Field Galaxy Survey (6dFGS) \citep{6dF}. The Baryon Oscillation Spectroscopic Survey (BOSS) \citep{Dawson12}, part of SDSS III \citep{Eis11}, built on this legacy to obtain the first percent level BAO measurements \citep{alph}. Results from the completed, Data Release (DR) 12, sample of BOSS galaxies were presented in \citet{Acacia}. 

As well as using galaxies as direct tracers of the BAO, analyses of the Lyman-$\alpha$ Forest in quasar spectra with BOSS have provided cosmological measurements at $z\sim2.3$ (e.g. \citealt{Delubac15,Bautista17}). However, between the current direct-tracer and Lyman-$\alpha$ measurements there is a lack of BAO measurements. Using quasars\footnote{In this work `quasar' is used as a synonym for quasi-stellar object (QSO) rather than quasi-stellar radio source; more specifically, we mean a high-redshift point source whose luminosity is presumably powered by a super-massive black hole at the centre of an unobserved galaxy.} as direct tracers of the density field offers the possibility of $1<z<2$ observations, with the main hindrance being their low space density and the difficulty of performing an efficient selection. The extended-Baryon Oscillation Spectroscopic Survey (eBOSS; \citealt{eboss}), part of SDSS-IV \citep{sdss4}, has been designed to target and measure redshifts for $\sim500,000$ quasars at $0.8<z<2.2$ (including spectroscopically-confirmed quasars already observed in SDSS-I/II). Although the space density will still be relatively low (compared with the densities of galaxies in BOSS, for example), eBOSS will offset this drawback by covering a significant fraction of the enormous volume of the Universe between redshifts 1 and 2.

Quasars were selected in eBOSS using two techniques. A ``CORE'' sample used a likelihood-based routine called XDQSOz to select from optical $ugriz$ imaging, combined with a mid-IR-optical colour-cut. An additional selection was made based on variability in multi-epoch imaging from the Palomar Transient Factory \citep{PTF}. These selections are presented in \citet{ebossQSO}, alongside the characterisation of the final sample, as determined by the early data. The early data were observed as part of SEQUELS (The Sloan Extended QUasar, ELG and LRG Survey, undertaken as part of SDSS-III and SDSS-IV; described in the appendix of \citealt{DR12}), which acted as a pilot survey for eBOSS. SEQUELS used a broader quasar selection algorithm than that adopted for eBOSS, and a subsampled version of SEQUELS forms part of the eBOSS sample.

In this paper we present BAO measurements obtained from eBOSS, using quasars from the DR14 dataset to measure the BAO distance to redshift 1.5. These measurements represent the first instance of using the auto-correlation of quasars to measure BAO and the first BAO distance measurements between $1 < z < 2$. The low space density of quasars means that reconstruction techniques \citep{Eis07rec} are not expected to be efficient, but we are still able to obtain a 4.4 per cent BAO distance measurement at greater than 2.5$\sigma$ significance. The results are an initial exploration of the power of the eBOSS quasar data set. We expect many forthcoming studies to further optimize these BAO measurements, measure structure growth, and probe the primordial conditions of the Universe.

This paper is structured as follows. In Section \ref{sec:data} we describe how eBOSS quasar candidates were `targeted' for follow-up spectroscopy, observed, and how redshifts were measured. In Section \ref{sec:catalog}, we describe how these data are used to create catalogs suitable for clustering measurements. Section \ref{sec:analysis}, presents our analysis techniques, including explanations of our fiducial cosmology, how we measure clustering statistics, how we model BAO in these clustering statistics, and how we assign likelihoods to parameters that we measure. In Section \ref{sec:mocks}, we describe the two techniques used to produce a total of 1400 simulated realizations of the DR14 quasar sample, i.e., `mocks'. Section \ref{sec:mocktests} reviews tests of our methodology using mocks; these tests allow us to define our methodological choices for combining measurements from different estimators. Section \ref{sec:results} presents the clustering of the DR14 quasar sample and the BAO measurements, including numerous robustness tests on these measurements. We then present an updated BAO distance ladder and place our measurement in the larger cosmological context in Section \ref{sec:cosmo}. We conclude with a preview of forthcoming cosmological tests expected to be performed using eBOSS quasar data and additional tracers in Section \ref{sec:con}.

%section
\section{Data}
\label{sec:data}
In this section, we review the imaging data that was used to define a sample of quasar candindate `targets' intended for spectroscopy. We then describe how we obtain spectroscopy for each target and then identify quasars and measure redshifts from this output. The process of transforming these data into large-scale structure (LSS) catalogs is described in Section \ref{sec:catalog}.

\subsection{Imaging}

All eBOSS quasar targets selected for LSS studies are selected on imaging from SDSS-I/II/III and the Wide Field Infrared Survey Explorer (WISE, \citealt{Wright2010}). We describe each dataset below.

SDSS-I/II \citep{York00} imaged approximately 7606 deg$^2$ of the Northern galactic cap (NGC) and approximately 600 deg$^2$ of the Southern galactic cap (SGC) in the $ugriz$ photometric pass bands \citep{F,Smith2002,Doi2010}; 
these data were released as part of the SDSS Data Release 7 (DR7, \citealt{DR7}). SDSS-III \citep{Eis11} obtained additional photometry in the SGC to increase the contiguous footprint to 3172 deg$^2$ of imaging in the SGC, released as part of DR8 \citep{DR8}, alongside a re-processing of all DR7 imaging. The astrometry of this data was subsequently improved in DR9 \citep{DR9}. All photometry was obtained using a drift-scanning mosaic CCD camera \citep{C} on the 2.5-meter Sloan Telescope \citep{Gunn06} at the Apache Point Observatory in New Mexico, USA. 

The eBOSS project does not add any imaging area to that released in DR8, but takes advantage of updated calibrations of that data. \cite{Schlafly2012} applied the ``uber-calibration" technique presented in \cite{Pad08} to Pan-STARRS imaging \citep{Kaiser2010}. This work resulted in an improved global photometric calibration with respect to SDSS DR8, which is internally applied to SDSS imaging. Residual systematic errors in calibration are reduced to sub per-cent level on all photometric bands \citep{Finkbeiner2016}, and poorly constrained zero-points are much improved. The photometry with updated calibrations was released with SDSS DR13 \citep{DR13}. 

The WISE satellite observed the entire sky using four infrared channels centred at 3.4 $\mu$m (W1), 4.6 $\mu$m (W2), 12 $\mu$m (W3) and 22 $\mu$m (W4). The eBOSS quasar sample uses the W1 and W2 bands for targeting; see \cite{ebossQSO} for details. All targeting is based on the publicly available {\it unWISE} coadded photometry force matched to SDSS photometry presented in \cite{Lang2014}.

\subsection{Spectroscopic Observations}

Quasar target selection for eBOSS is described in \cite{ebossQSO}. Objects that satisfy the target selection and which do not have a previously known and secure redshift are flagged as ${\tt QSO\_EBOSS\_CORE}$ and assigned optical fibers (via a process termed {\it tiling} - see Section~\ref{sec:mask}), and selected for spectroscopic observation. Spectroscopy is collected using the BOSS double-armed spectrographs \citep{Smee13}, covering the wavelength range 3600 to 10000 \AA\ with R$=1500-2600$. In BOSS, the pipelines to process data from CCD-level to 1d spectrum level to redshift are described in \cite{DR13} and \cite{Bolton12}. 

We divide the sources of secure redshift measurement into three classes:
\begin{itemize}
\item Legacy: these are quasar redshifts obtained by SDSS I/II/III via non-eBOSS related progams;
\item SEQUELS: these are quasar redshifts obtained from the Sloan Extended QUasar, ELG and LRG Survey (SEQUELS) (designed as pilot survey for eBOSS, again see \citealt{ebossQSO});
\item eBOSS: these are previously unknown quasar redshifts obtained by the eBOSS project.
\end{itemize}

\subsubsection{Legacy}
The eBOSS program does not allocate fibers to targets previously observed that have a confident spectroscopic classification and a reliable redshift from previous SDSS observations. A target is considered to have a ``confident'' classification if neither ${\tt LITTLE\_COVERAGE}$ nor ${\tt UNPLUGGED}$ are flagged in the ${\tt ZWARNING}$ bitmask. A target is considered to have a ``good" redshift if it is not labeled ${\tt QSO ?}$ or ${\tt QSO\_Z ?}$ in the DR12 quasar catalog of \cite{Paris2017}. These targets, collectively termed legacy, typically have good, visually inspected, redshifts collated from SDSS-I, II and III data. Redshifts acquired before BOSS are obtained from a combination of \cite{Schneider2010} and a catalogue of known stellar spectra from SDSS-I/II. Targets observed during BOSS that resulted in a confident spectral classification and redshift are documented in the DR12 quasar catalogue (DR12Q, \citealt{Paris2017}), and are not re-observed in eBOSS. These known objects are therefore flagged as either ${\tt QSO\_BOSS\_TARGET}$, ${\tt QSO\_SDSS\_TARGET}$ or ${\tt QSO\_KNOWN}$ (see section 4.4 of \citealt{ebossQSO} for full details on how these flags are set). Targets that were previously observed in SDSS-I/II/III but failed to result in a confident classification (i.e., had at least one of ${\tt LITTLE\_COVERAGE}$ or ${\tt UNPLUGGED}$ set) or a good redshift determination (i.e., were not labeled ${\tt QSO?}$ or ${\tt QSO\_Z?}$ in DR12Q) were targeted for re-observation by either SEQUELS or eBOSS.

\subsubsection{SEQUELS}
SEQUELS is a spectroscopic program started during SDSS-III that was designed as a pilot survey for eBOSS. The total program consists of 117 plates, 66 of which were observed during BOSS and are included in DR12 \citep{DR12}. The remaining 51 plates were observed during the 1st year of the eBOSS program and were released in DR13 \citep{DR13}. The target selection for SEQUELS is by construction deeper and less constrained than the finalized eBOSS target selections, so only a (large) fraction of the SEQUELS targets satisfy the eBOSS final selection criteria.\footnote{See \S5.1 of Myers et al. (2015) for full details of the minor selection differences between SEQUELS and the rest of eBOSS.} The SEQUELS area is not re-observed in eBOSS and, for the purpose of these catalogues, we treat SEQUELS targets that pass the final eBOSS target selection in an identical manner to eBOSS targets in the eBOSS footprint.

EBOSS\_TARGET0 holds the targeting flags for SEQUELS targets, whereas EBOSS\_TARGET1 contains the targeting flags for eBOSS targets. In the target collate file and in the LSS catalogues, {\em all} targets that pass the eBOSS selection have the appropriate EBOSS\_TARGET1 set, irrespective of whether they lie in the SEQUELS or eBOSS footprint. These flags match those that will exist in the publicly released catalogs.

\subsubsection{eBOSS}

The eBOSS project, naturally, represents the bulk of our observations --- over 75 per-cent of new redshifts in the DR14 LSS catalogues were observed during the eBOSS program. The target selection algorithm for quasars includes both LSS and Lyman$\alpha$ quasar targets. We use only the LSS quasars, which have the QSO\_CORE bit set in the targeting flags. The DR14 sample includes two years of eBOSS observations.

\subsection{Measuring Redshifts}

Robust spectral classification and redshift estimation is a challenging problem for quasars. In particular, the number and complexity of physical processes that can affect the spectrum of a quasar make it difficult to precisely and accurately disentangle systemic redshift (i.e., as a meaningful indicator of distance) from measured redshift (e.g., \citealt{Hewett2010}). SEQUELS observations taken during SDSS-III (representing around half of the SEQUELS program) were all visually inspected, and helped define our process for identifying quasar candidates. As detailed in \cite{eboss}, 91 per cent of quasar spectra targeted for clustering studies are securely classified with an automated pipeline (according to said pipeline) and less than 0.5 per cent of these classifications were found to be false when visually examined. The automated classification fails to report a secure classification in the remaining nine per cent of cases and these are visually inspected, which is able to identify approximately half of these as quasars.

Information on all eBOSS quasars is detailed in the DR14 quasar catalogue (DR14Q, \citealt{Paris2017b}), the successor to DR12Q, with the important distinction that the vast majority of LSS quasars are not visually inspected. DR14Q combines the LSS pipeline and visual inspection results together and provides a variety of value-added information. In particular, it contains three automated estimates of redshift that we consider in our LSS catalogs:
\begin{itemize}
\item The SDSS quasar pipeline redshifts, denoted `Z$_{\rm PL}$', and documented in \cite{Bolton12}. The pipeline uses a PCA decomposition of galaxy and quasar templates, alongside a library of stellar templates, to fit a linear combination of four eigenspectra to each observed spectrum. 
\item A redshift estimate based on the location of the maximum of the MgII emission line blend at $\lambda = 2799$\AA\ , denoted `Z$_{\rm MgII}$'.  The MgII broad emission line is less susceptible to systematic shifts due to astrophysical effects and, when a robust measurement of this emission line is present, it offers a minimally-biased estimate of a quasar's systemic redshift (see e.g. \citealt{Hewett2010, Shen2016}).
\item A `Z$_{\rm PCA}$'  estimate, as documented in \cite{Paris2017}. `Z$_{\rm PCA}$' uses a PCA decomposition of a sample of quasars with redshifts measured at the location of the maximum of the MgII emission line, and fits a linear combination of four eigenvectors to each spectrum.
\end{itemize}

Whereas Z$_{\rm MgII}$ offers the least biased estimate of the quasar's systemic redshift, it is more susceptible to variations in S/N, and therefore Z$_{\rm PCA}$ is able to obtain the accuracy of Z$_{\rm MgII}$ with increased robustness provided by utilizing the information from the full spectrum (see Fig. 10 of \citealt{eboss}). 

DR14Q also contains a redshift, `Z', which it considers to be the most robust of the available options, in that these redshifts are known to have the lowest rate of catastrophic failures (and can be any of the three options above, depending on the particular object). Further details will be available in \cite{Paris2017b}. We will test the robustness of our results to the redshift estimates by also testing BAO measurements where we use Z$_{\rm PCA}$ as the redshift in all cases where it is available. Further tests, especially those focusing on the impact on redshift-space distortion (RSD) measurements, will be presented in \cite{Zarrouk17}.

\begin{figure}
\includegraphics[width=84mm]{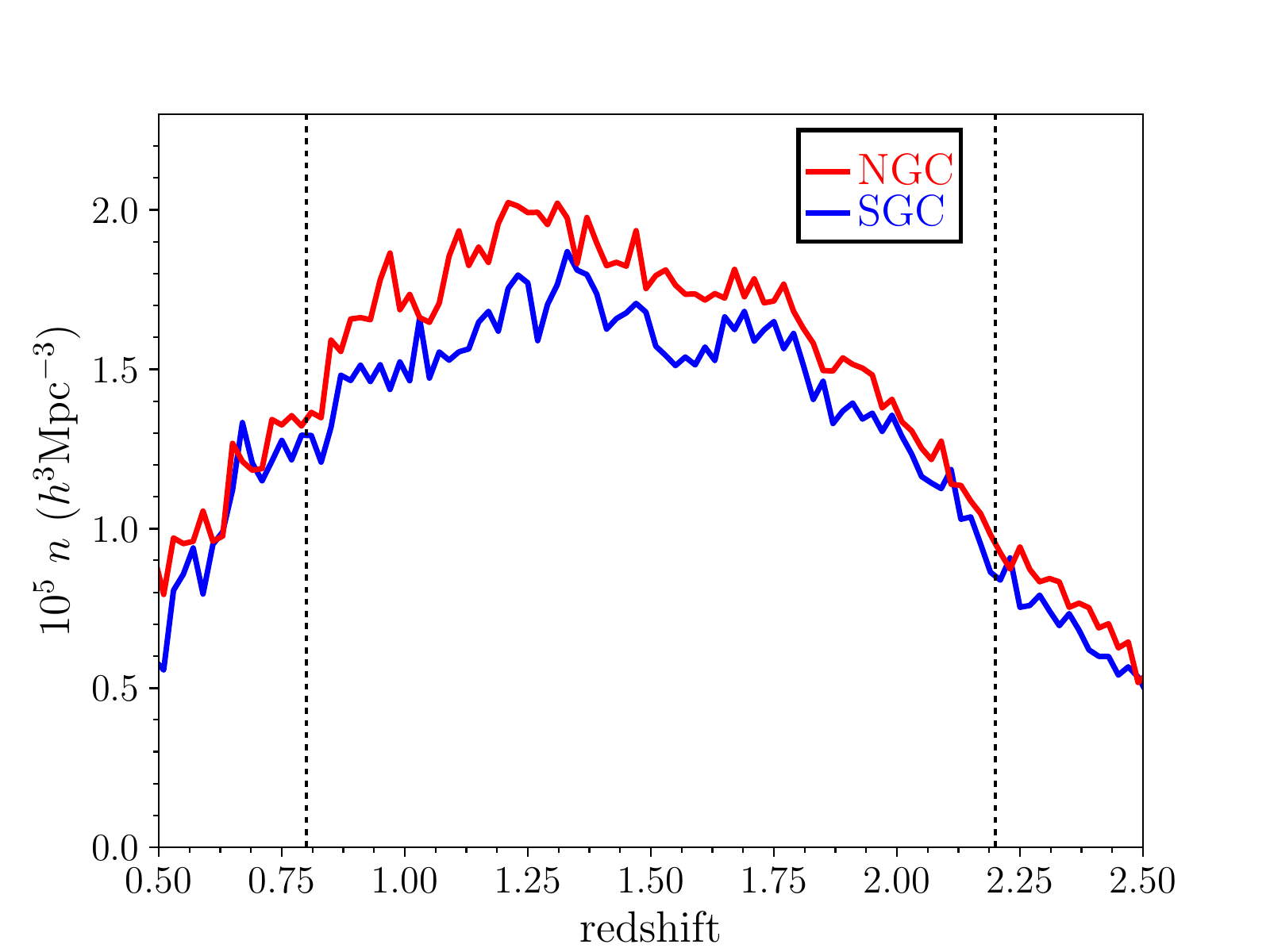}
  \caption{The redshift distribution of the DR14 quasar sample, for 111,633 quasars in the NGC and 75,887 in the SGC. We use the data with $0.8 < z < 2.2$ for clustering statistics; this redshift region is marked with dotted lines. The $n(z)$ is slightly different in the NGC and SGC, due to known differences in the targeting efficiency, and we thus treat the two regions separately.} 
  \label{fig:nz}
\end{figure}

The redshift distribution of the DR14 LSS quasar sample is displayed in Fig. \ref{fig:nz}. The curves show the result for the fiducial redshift sample. Our study uses the data with $0.8 < z < 2.2$. The target sample selection was optimized to yield quasars with $0.9 < z < 2.2$ \citep{ebossQSO}. At lower redshifts, morphological cuts affect the sample selection; at higher redshift the redshift measurement is less secure. We can securely select quasars to $z<0.8$, but given that BAO at lower redshifts is better sampled by galaxies, we impose the $z>0.8$ cut. Affecting our choice of a high-redshift cut is that quasars with $z > 2.2$ are used for Ly-$\alpha$ clustering measurements and we wish to cleanly separate the two volumes used for BAO measurements\footnote{We are likely to re-evaluate this choice in future studies.}. The data in the NGC (red) has a slightly greater number density than that of the SGC (blue). The imaging properties in the two regions are somewhat different, and, as explained in \cite{ebossQSO}, we expect a more efficient target selection (and thus yield of successful quasar redshifts) in the NGC. We describe weights that are applied to correct for the variations in targeting efficiency in Section \ref{sec:weights}.

\section{LSS Catalogs}
\label{sec:catalog}

In this section, we detail how the quasar target and redshift information is combined to create LSS catalogs suitable for large-scale clustering measurements.\footnote{The catalogues will be available at this website \url{https://data.sdss.org/sas/dr14/eboss/lss/}, after eBOSS DR14 studies are complete.} The clustering of the eBOSS quasar sample has already been studied by \cite{Rod16} and \cite{Laurent16}. LSS catalogs were generated for these studies in similar fashion to the methods we outline below, which are closely matched to the methods described in \cite{Reid16}. In particular, \cite{Rod16,Laurent16} found that the clustering amplitude of the sample, and its redshift evolution, is consistent with the assumptions used in \cite{Zhao16forecast} and that the clustering can be modeled with the type of simulation techniques that have been successfully applied to galaxy samples. 

\subsection{Footprint}
\label{sec:mask}

Targets that pass the target selection algorithm, and for which there are no known good redshifts, are fed into a tiling algorithm \citep{Blanton2003}, that allocates spectroscopic fibers to targets within a $3^{\circ}$ tile. Allocation is done in a way that maximises the number of fibers placed on targets, considering the constraints imposed by a pre-set target priority list and the $62"$ exclusion radius around each fiber \citep{eboss}. The algorithm is sensitive to the target density on the sky, so overdense regions tend to be covered by more than one tile. This overlap of tiles locally resolves some collision conflicts, but all others are dealt with separately. For eBOSS and SEQUELS, quasars can have collisions with fellow quasars or higher priority target classes. Collisions with other target classes (including Lyman$\alpha$) are simply deemed `missed' observations and will be treated as random. Collisions with fellow quasars are termed {\it fiber collisions} or {\it close pair collisions}; see Section~\ref{sec:weights_c}. Approximately 40\% of the eBOSS area is covered by more than one tile. 

We use the MANGLE software package \citep{Swanson2008} to decompose the sky into a unique set of {\it sectors}, within each we compute a survey completeness.  Within each sector we define the following:

\begin{itemize}
\item N$_{\rm legacy}$: the number targets with previously known redshifts (excluded from tiling);
\item N$_{\rm good}$: the number of fibers that yield good quasar redshifts;
\item N$_{\rm zfail}$: the number of fibers from which a redshift could not be measured;
\item N$_{\rm badclass}$: the number of targets with spectroscopic classification that does not match its target class, for our quasar sample, these are exclusively galaxies\footnote{While we exclude them from our analysis, these are likely an interesting sample of object.};
\item N$_{\rm cp}$: the number of targets which did not receive a fiber due to being in a collision group (or ``close-pair");
\item N$_{\rm star}$: the number of spectroscopically confirmed stars;
\item N$_{\rm missed}$: the number of quasar targets to be observed in the future or not observed because of a collision with a different eBOSS target class.
\end{itemize}

\begin{table}
\centering
\caption{Basic properties of the quasar LSS catalogues. The quantities are summed over all sectors, with no redshift cuts. $\rm N_{Q} = N_{good} + N_{legacy}$. $\rm N_{eff}$ is the effective total number of quasars, after correcting for redshift failures and fibre-collisions: $\rm N_{eff} = \sum (w_{cp} + w_{zfail} - 1)$. {\it Unweighted area} is the sum of the area of all sectors with $\rm C_{eBOSS} > 0.5$; {\it weighted area} multiplies this area by the completeness in each sector and {\em weighted area post-veto} multiplies this area by the total fraction of vetoed area. All other quantities are defined in the text. }
\begin{tabular}{lrrr}
\hline
	& NGC & SGC & Total \\\hline\hline
$\rm \bar{N}_{QSO}$ & 116866 & 77935& 194801 \\
$\rm \bar{N}_{good}$ & 78425 & 58277& 136702 \\
$\rm \bar{N}_{legacy}$ & 38441 & 19658& 58099 \\
$\rm \bar{N}_{zfail}$ & 3598 & 2865& 6463 \\
$\rm \bar{N}_{cp}$ & 3126 & 2352& 5478 \\
$\rm \bar{N}_{badclass}$ & 8908 & 5564& 14472 \\
$\rm \bar{N}_{star}$ & 3782 & 4517& 8299 \\
$\rm \bar{N}_{eff}$ & 123903 & 82876& 206779 \\
Unweighted area (deg$^2$) & 1356 & 1035  & 2391 \\
Weighted area (deg$^2$) & 1288 & 995  & 2283 \\
Weighted area post-veto (deg$^2$) & 1215 & 898 & 2113\\
\hline\hline
\label{tab:catalogues}
\end{tabular}
\end{table}

A summary of the above numbers in each of our target samples, summed over all sectors, is given in Table \ref{tab:catalogues}. We define a targeting completeness per sector and per target class as

\begin{equation}
C_{\rm eBOSS} = \frac{\rm N_{\rm good} + N_{\rm zfail} +  N_{\rm badclass} + N_{\rm cp} + N_{\rm star} }{\rm N_{\rm good} + N_{\rm zfail} +  N_{\rm badclass} + N_{\rm cp} + N_{\rm star}  +N_{\rm missed}}.
\label{eq:compspec}
\end{equation}

\noindent Thus, $C_{\rm eBOSS}$ tracks the fiber-allocation completeness of the eBOSS spectroscopic observations. This is the completeness that defines the eBOSS mask and which will be later used to construct random catalogues with a matched on-sky completeness (see Section~\ref{sec:randoms}). Inspection of Eq. \ref{eq:compspec} reveals that $C_{\rm eBOSS}$ is impacted only by  $N_{\rm missed}$; objects that were not assigned a fiber due to a fiber collision are treated separately (see Section \ref{sec:weights_c}).

\begin{figure}
\includegraphics[scale=0.3]{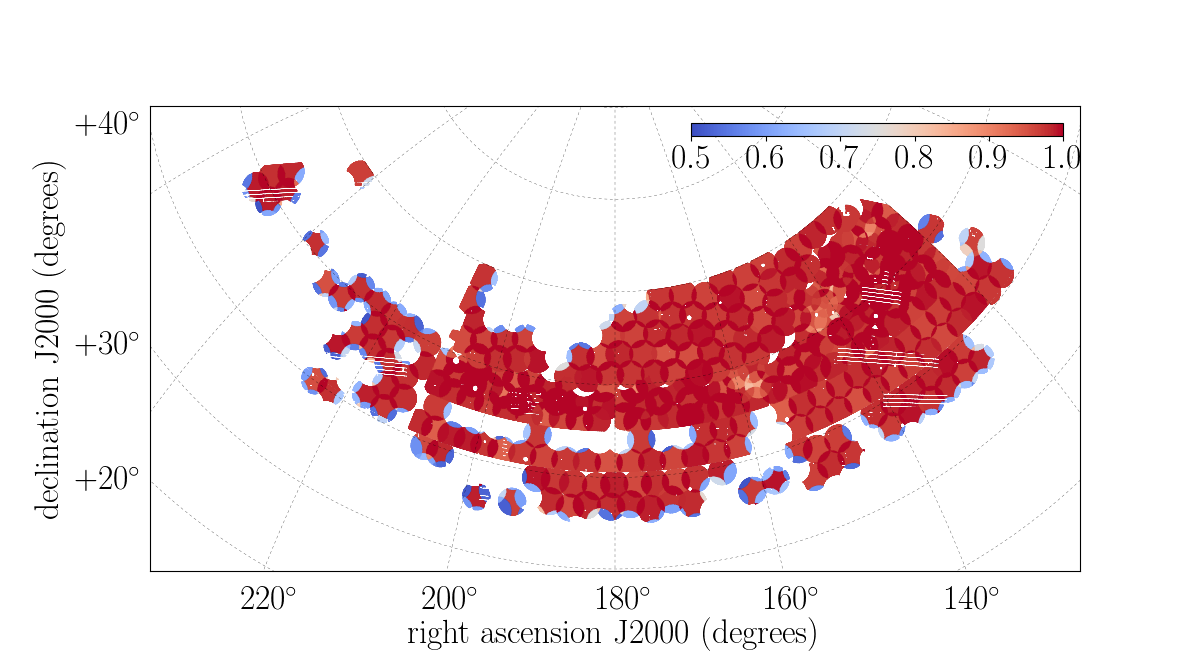}
\includegraphics[scale=0.3]{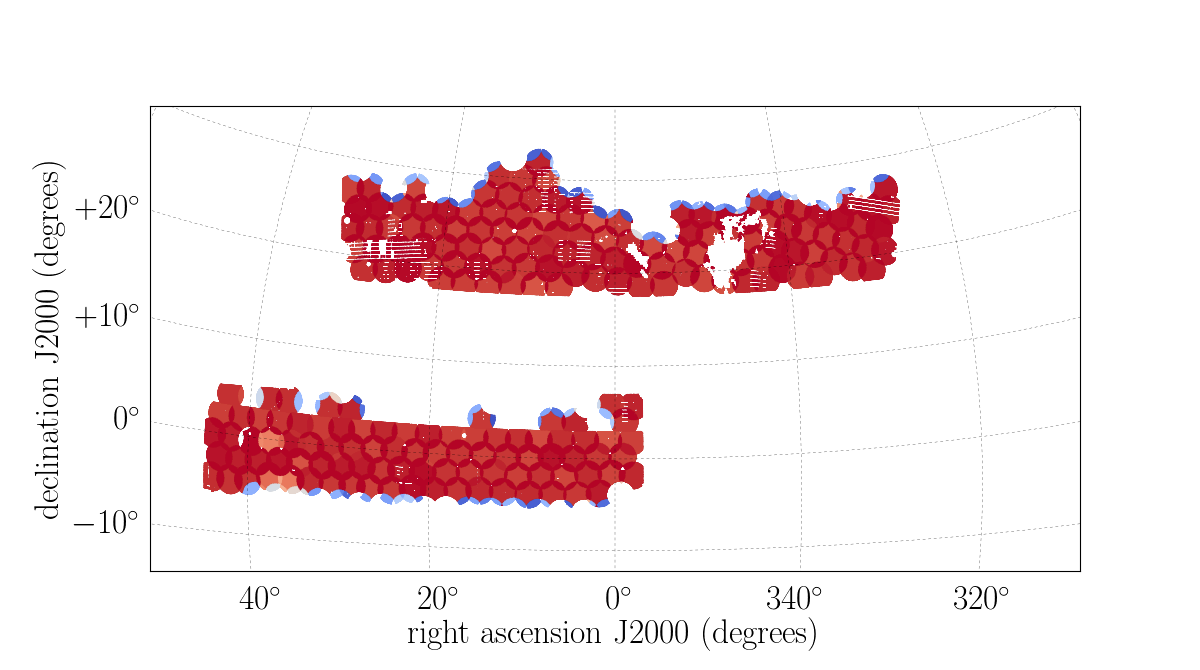}
\caption{The footprint of the eBOSS DR14 quasar sample. The top panel displays the portion of the footprint in the NGC and the bottom panel the SGC. The colour mapping indicates the observational completeness, C$_{\rm eBOSS}$, as defined in the text. }
\label{fig:footprint_QSO}
\end{figure}

Legacy targets are 100\% complete, since they have already been observed. In order to account for this, we follow the same procedure as in BOSS \citep{Reid16} and sub-sample legacy targets to match the value of C$_{\rm eBOSS}$ in each sector. SEQUELS and eBOSS observations are very similar and thus we treat them the same way, without distinction in the LSS catalogues. We keep all sectors with C$_{\rm eBOSS} > 0.5$ in the LSS catalogues; the average completeness of the remaining sectors is high, averaging 95 and 96 per-cent in the North and South Galactic caps, respectively. The footprint of the DR14 LSS catalogues, coloured by the value of $C_{\rm eBOSS}$ in each sector, is shown in Fig.~\ref{fig:footprint_QSO}. The completeness is generally quite high, except around the edges of the footprint where future observations will overlap with the DR14 data. Veto masks have also been applied and are detailed in the following subsection.

Additionally, we define a redshift completeness per sector as

\begin{equation}
C_{\rm z} = \frac{\rm N_{good}}{\rm N_{good} + N_{fail}},
\end{equation}

\noindent which tracks the quasar redshift efficiency averaged over each sector. We use this completeness value only to remove sectors with C$_{\rm z} < 0.5$. Redshift failures themselves are corrected for separately (see Section \ref{sec:weights_c}).

\subsection{Veto Masks}\label{sec:veto_masks}

A number of veto masks are used to exclude sectors in problematic areas. For the DR14 quasar sample, we apply the same veto masks as in BOSS DR12 \citep{Reid16}, removing regions due to:

\begin{itemize}
\item Bad photometric fields, including cuts on seeing and Galactic extinction. In total, this mask excludes approximately 5 per-cent of the area. Cuts on extinction and seeing are only significant in the SGC (3.2 per-cent of the SGC area is excluded by the seeing cut and 2.6 per-cent by the extinction cut).
\item Bright stars, based on the Tycho catalog (\citealt{Tycho}; excluding 1.8 per cent of the area); 
\item Bright objects, including, e.g., stars not in the Tycho catalog and bright galaxies (\citealt{Rykoff14}; excluding 0.05 per cent of the area);
\item Centerposts, which anchor the spectrographic plates and prevent any fibers from being placed there (excluding $< 0.01$ per cent of the area);
\end{itemize}
We use the \cite{SFD} map to determine extinction values and we remove areas with $E(B-V)>0.15$. For seeing, we use the value labeled `PSF\_FHWM' in the catalogues and remove areas where is greater than 2.3, 2.1, 2.0 in the g, r, and i band, respectively. Further details on these masks and their motivation can be found in section 5.1.1 of \cite{Reid16}. The veto masks have been applied to Fig. \ref{fig:footprint_QSO}. The large gap in coverage in the SGC at RA$\sim345^{\rm o}$, Dec$\sim 22^{\rm o}$ is due to the extinction mask. The horizontal striped patterns are due to the photometric bad fields or poor seeing in the SDSS imaging. The other veto masks are generally too small to be distinguishable.

\subsection{Spectroscopic Completion Weights}
\label{sec:weights_c}
The spectroscopic completeness of the sample is affected by multiple factors. Again, our process for accounting for this incompleteness matches that described in \cite{Reid16}. The simplest is that not all targets in a given sector have been observed. We account for this effect by down-sampling the random catalogs by the completeness fraction. Targets also lack redshifts due to fiber collisions and redshift failures, and we describe how these are treated below.

Not all observations yield a valid redshift. Redshift failures do not happen randomly on a tile (see e.g. \citealt{Laurent16}), meaning they cannot be accounted for uniformly within a sector. Instead, as in previous BOSS analyses (e.g. \citealt{Reid16}), we choose to transfer the weight of the lost target to the nearest neighbour with a good redshift and spectroscopic classification in its target class, within a sector (this can be a quasar, a star or a galaxy, provided it is targeted as a quasar). This weight is tracked by WEIGHT\_NOZ ($w_{\rm noz}$) in the LSS catalogues. WEIGHT\_NOZ is set to 1 by default for all objects, and incremented by $+1$ for objects with a neighbouring redshift failure. The median separation between a redshift failure and its up-weighted neighbour is 0.06$^{\rm o}$. This corrective scheme assumes that the redshift distribution of the redshift failures is the same as that of the good redshifts. This is not expected to be strictly true, but assumed for simplicity, given the small number of targets that are corrected as redshift failures (approximately 3.4 per-cent in the NGC and 3.6 per-cent in the SGC). This concern does not affect out BAO analysis, but its impact on RSD analysis is currently being studied.  

Targets missed due to fiber collisions do not happen randomly on the sky - they are more likely to occur in overdense regions. Targets lost to fiber collisions have therefore a higher bias than average, and we must apply a correction to account for this. We correct for these fiber collisions by transferring the weight of the lost target to the nearest neighbour of the same target class with a valid redshift and spectroscopic classification. This weight is tracked by WEIGHT\_CP ($w_{\rm cp}$) in the LSS catalogues. Legacy targets are allowed to accrue close-pair correction weights, and legacy targets are downsampled such that the number of eBOSS-legacy close-pairs matches the number of close-pairs in eBOSS within each sector. Like WEIGHT\_NOZ, it is set to 1 as default for all objects and incremented by $+1$ for every neighbouring fiber collision. A total of 4.0 per cent of the eBOSS quasar targets are corrected as close-pairs in the NGC; this fraction is 3.0 per cent in the SGC.

Redshift failures are allowed to accrue weight from neighbouring close pairs, in which case the closest neighbour sees $w_{\rm noz}$ incremented by the total $w_{\rm cp}$ of redshift failure. For example, it is possible for a quasar target to be unobserved due to a collision with another quasar target. The observed quasar target is given $w_{\rm cp}=2$, but we then fail to obtain a good redshift. The nearest neighbour to this observed quasar target is thus given $w_{\rm noz} = 3$.

Thus, each quasar is given a spectroscopic completeness weight, $w_{\rm c}=(w_{\rm cp}+w_{\rm noz}-1)$ to be used for any counting statistics.

\subsection{Systematic Weights for Dependencies on Imaging Properties}
\label{sec:weights}
\begin{figure}
\includegraphics[width=84mm]{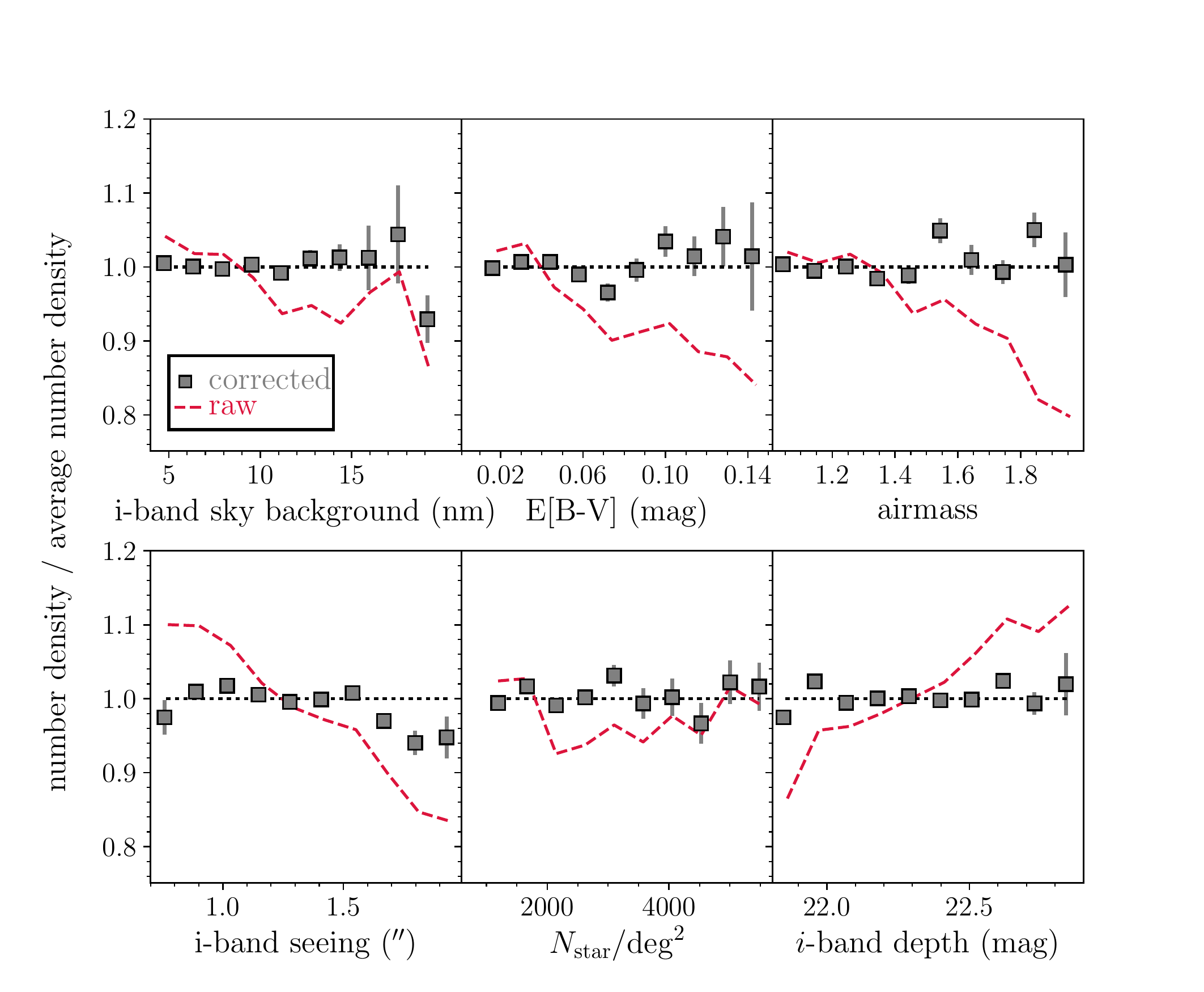}
  \caption{The relationship between the number density of the DR14 quasar sample and various potential systematics before (dashed crimson curves, labeled `raw') and after (grey squares, labeled `corrected') weighting for limiting magnitude (depth) and Galactic extinction (E[B-V]). Weighting for limiting magnitude and E[B-V] removes correlations with other potential systematic quantities. 
  }
  \label{fig:sysall}
\end{figure}

As described in \cite{Laurent16}, weights are required for the DR14 quasar sample in order to remove spurious dependency on the 5$\sigma$ limiting magnitude (`depth') and Galactic extinction. Quasars are more securely identified where the depth is best and Galactic extinction is the variable that we find most affects differences in depth between the SDSS imaging bands, as they were nearly simultaneously observed.

For the DR14 quasar sample, we define weights based on the depth in the $g$-band, in magnitudes (including the effect of Galactic extinction on this depth), and the Galactic extinction in units $E(B-V)$, using the map determined by \cite{SFD}. These are the important observational systematics identified in \cite{Laurent16}. We define the weights based on the sample DR14 quasars with $0.8 < z < 2.2$ (already passed through the steps defined in the preceding section). Compared to \cite{Laurent16}, our results differ in that we use the full DR14 set (approximately doubling the sample size) to determine the weights and that we define the weights separately for the NGC and SGC. As in \cite{Ross12,Ross17} and \cite{Laurent16}, we define the weights based on fits to linear relationships. We first determine the dependency with depth and then with extinction, after applying the weights for depth. The total weight is the multiplication of the two weights. Thus
\begin{equation}
w_{\rm sys} = \frac{1}{(A_d+dB_d)(A_e+eB_e)},
\end{equation}
where $d$ is the $g$ band depth (in magnitudes) and $e$ is the Galactic extinction (in $E(B-V)$).
The best-fit coefficients are $A_d = -3.52, B_d = 0.195, A_e = 1.045, B_e = -2.01$ for the NGC and $A_d = -6.20, B_d = 0.31, A_e = 1.052, B_e = -1.00$ for the SGC. The differences in the coefficients for the two regions make it clear it is necessary to separate them for analysis of the DR14 sample. Fig. \ref{fig:sysall} presents the relationship between the projected number density quasars and potential systematic quantities, combining the NGC and SGC. After weighting for depth and Galactic extinction (red squares) the systematic trends are removed.

\begin{figure}
\includegraphics[width=84mm]{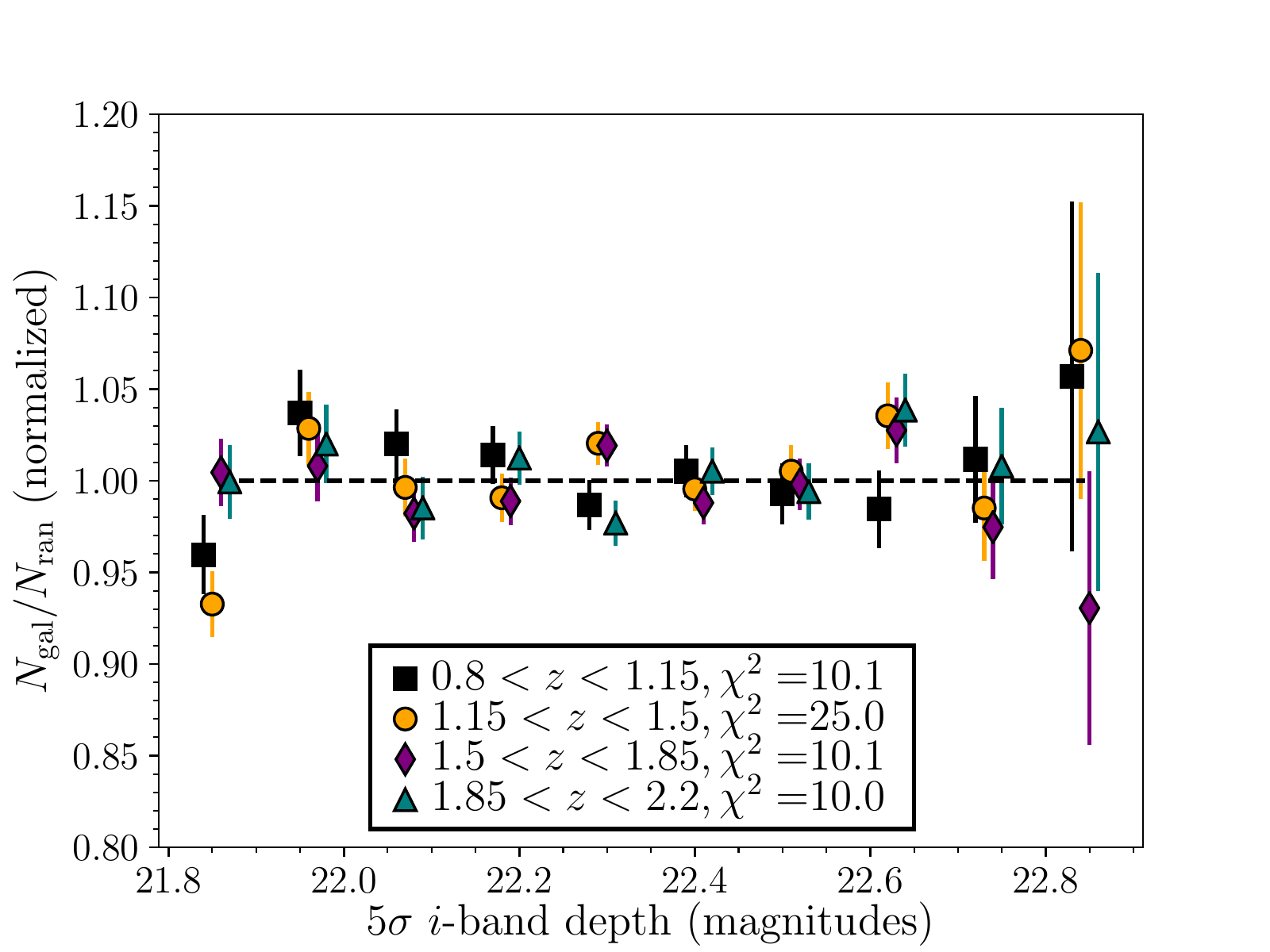}
  \caption{The relationship between the number density of the DR14 quasar sample and the $i$-band 5$\sigma$ limiting magnitude (`depth') for four slices in redshift, after weights for depth and Galactic extinction have been applied. No systematic trends with redshift are apparent. 
  }
  \label{fig:depthvz}
\end{figure}

Fig. \ref{fig:depthvz} displays the relationship between quasar density and the depth when dividing the sample into four redshift bins. No systematic trends are apparent with redshift, suggesting that the systematic relationships do not need to be defined as a function of the colour/magnitude of the quasars. The $\chi^2$ for the null test for the quasars with $1.15 < z < 1.5$ is large --- 25 for 10 degrees of freedom --- but this result is dominated by a single 4$\sigma$ outlier at the worst depth. For the 9 bins at greater depth, the $\chi^2$ is 12. We will demonstrate that our results are robust to any fluctuations in density imparted by the depth fluctuations.

\subsection{Random Catalogues} \label{sec:randoms}

Random catalogues are constructed that match the angular and radial windows of the data, but with approximately 40 times the number density. Such catalogs are required for both correlation function and power spectrum estimates of the clustering of the DR14 quasar sample, as detailed in Section \ref{sec:analysis}.

We begin by using the MANGLE software to generate a set of points randomly distributed in the eBOSS footprint, where the angular number density in each sector is subsampled to match the value of C$_{\rm eBOSS}$ in that sector. We then run the random points through the same veto masks that are applied to the data (see Section~\ref{sec:veto_masks}). Finally, we assign each random point a redshift that is drawn from the distribution of data redshifts that clear the veto mask. The draws are weighted by the total quasar weight given by w$_{\rm tot}=w_{\rm sys}*w_{\rm c}$, such that the weighted redshift distribution of data and randoms match.

\section{Methodology}
\label{sec:analysis}

%subsection
\subsection{Fiducial Cosmology}
\label{sec:fidcosmo}

We use a flat, $\Lambda$CDM cosmology with $\Omega_m=0.31$, $\Omega_{b}h^2=0.022$, $\sum m_\nu=0.06\ $eV and $h=0.676$, where the subscripts $m$, $b$ and $\nu$ stand for matter, baryon and neutrino, respectively, and $h$ is the standard dimensionless Hubble parameter. These choices match the fiducial cosmology adopted for BOSS DR12 analyses \citep{Acacia}. One set of mocks we use, the EZmocks (see Section \ref{sec:ezmock}), use the cosmology of MultiDark-PATCHY \citep{PATCHY14,Kitaura15} used in previous BOSS analyses. The other set of mocks we use, QPM mocks (see Section \ref{sec:QPM}), uses a geometry that matches our fiducial cosmology but with $\Omega_\nu=0$. The properties of the cosmologies we use are listed in Table \ref{tab:cosmo}.
Following the values provided in Table \ref{tab:cosmo}, the BAO distance parameter at the effective redshift of the quasar sample, $D_V(z_{\rm eff})$ with $z_{\rm eff}=1.52$  (see Eq. \ref{DV_definition} for definition), are $3871.0\,{\rm Mpc}$
 for both the fiducial cosmology and for the QPM cosmology and $3871.7\,{\rm Mpc}$ for the EZ mocks cosmology. A separate factor entering our analysis is the value for the comoving sound horizon at the baryon drag epoch, $r_{\rm d}$; this parameter sets the position of the BAO scale in our theoretical templates. The different cosmologies and lack of a neutrino mass in the mocks shift $r_{\rm d}$ to be less than the fiducial by just over $0.1\,{\rm Mpc}$ for each type of mock we use.

\begin{table}
\centering
\caption{Fiducial cosmology used in our BAO analysis and true cosmology for the QPM and EZ mocks, described in Section \ref{sec:mocks}. Given that the cosmology in which we analyse the mocks is slightly different than their own one, we expect a shift in the BAO peak position with respect to the fiducial position, $\alpha$ (see Eq. \ref{alpha_definition} for definition). We also provide the values for the comoving sound horizon at the baryon drag epoch, $r_{\rm d}$. The exact values used for the EZ mock are $\Omega_{m} = 0.307115$ and $h=0.6777$, which have been rounded to three significant figures below}
\begin{tabular}{lcccccc}
\hline
\hline
case & $\Omega_{m}$ & $h$ & $\Omega_bh^2 $ & $\sum m_\nu$  & $\alpha$ & $r_{\rm d}$ (Mpc)\\
\hline
fiducial & 0.31 & 0.676 & 0.022 & 0.06 eV & - & 147.78\\
QPM & 0.31 & 0.676 & 0.022 & 0 & 1.00108 & 147.62\\
EZ &  0.307 & 0.678 &  0.02214 & 0 & 1.00101 & 147.66\\
\hline
\label{tab:cosmo}
\end{tabular}
\end{table}

%subsection
\subsection{Clustering Estimators}
We perform two complementary BAO analyses: i) in configuration space, where the observable is the angle (with respect to the line of sight) average (the monopole) of the correlation function; ii) in Fourier Space, where the observable is the monopole of the power spectrum. When the entire spectrum of frequencies and positions is considered, both correlation function and power spectrum contain identical information as one represents the Fourier transform of the other. However, since our spectral range is finite, the correlation function and power spectrum do not contain exactly the same information, although we expect a high correlation between results using either statistic. Long wavelengths are limited by the size of the survey and small wave-lengths are limited by the resolution of the analysis. Furthermore, we expect that any potential uncorrected observational or modelling systematics will affect  the correlation function and power spectrum differently. Thus, by performing two complementary analyses and combining them we expect to produce a more robust final result. 

For both analyses, we require the {\it data catalogue}, which contains the distribution of quasars (which can be an actual or synthetic distribution) and the {\it random catalogue}, which consists of a Poisson-sampled  distribution with  the  same  mask and  selection function as the data catalogue with no other cosmological correlations.
We count each data and random object as a product of weights. For the data catalogue, the total weight corrects for systematic dependencies in the imaging, $w_{\rm sys}$ and spectroscopic data, $w_{\rm c}$ (see Section \ref{sec:weights} and \ref{sec:weights_c}, respectively) multiplied by a weight, $w_{\rm FKP}$, that is meant to optimally ponderate the contribution of objects based on their number density at different redshifts. Conversely, for random catalogue, objects are weighted only by $w_{\rm FKP}$. The $w_{\rm FKP}$ weight is based on \cite{FKP} and defined as,
\begin{equation}
w_{\rm FKP}(z) = 1/[1+n(z)P_0],
\label{eq:wfkp}
\end{equation}
where $P_0$ is the amplitude of the power spectrum at the $k$ scale at which the FKP-weights optimise the measurement. For the expected BAO signature in the DR14 quasar sample this is $k\sim 0.14h$Mpc$^{-1}$ \citep{FB14}, and therefore, we use $P_0 = 6\times10^3\,[{\rm Mpc}\,h^{-1}]^{3}$. The FKP weights have only a small effect on our results, as the number density is both low and nearly constant, so the value of the weight varies by less than 10 per cent.

The total weight applied to each quasar is thus
\begin{equation}
w_{\rm tot} = w_{\rm FKP}w_{\rm sys}(w_{\rm cp}+w_{\rm noz}-1),
\label{eq:weight}
\end{equation}
while for each random object, the weight is simply $w_{\rm FKP}$.

%subsection
\subsubsection{Configuration Space}
For the  configuration space analysis the procedure we follow is the same as in \cite{alph}, except that our fiducial bin-size is 8 $h^{-1}$Mpc. We repeat some of the details here. We determine the multipoles of the correlation function, $\xi_{\ell}(s)$, by finding the redshift-space separation, $s$, of pairs of quasars and randoms, in units $h^{-1}$Mpc assuming our fiducial cosmology, and cosine of the angle of the pair to the line-of-sight, $\mu$, and employing the standard \cite{LS} method 
\begin{equation}
\xi(s,\mu) =\frac{DD(s,\mu)-2DR(s,\mu)+RR(s,\mu)}{RR(s,\mu)}, 
\label{eq:xicalc}
\end{equation}
where $D$ represents the quasar sample and $R$ represents the uniform random sample that simulates the selection function of the quasars. $DD(s,\mu)$ thus represent the number of pairs of quasars with separation $s$ and orientation $\mu$. In order to minimize any noise coming from the finite size of the random catalog, the random catalogs are many times the size of the data catalogs and the resulting counts are normalized accordingly. For the DR14 data, we use a random sample that is 40$\times$ as large as the data, which we have found is sufficiently large for our results to have converged within their quoted precision. For the mocks, we use larger random samples, 100$\times$ for the EZmocks and 70$\times$ for the QPM mocks. This is due to the fact that we use a single random catalog for all mocks. This eliminates any noise in the covariance matrix we determine from the mocks due to the finite size of the random catalogs. However, in order to obtain results to the precision expected for the full ensemble of mocks (e.g., to test their mean results) we require a random sample many times larger than required for a single realization.

We calculate $\xi(s,|\mu|)$ in evenly-spaced bins\footnote{The pair-counts are tabulated using a bin width of 1 $h^{-1}$Mpc and summed into x $h^{-1}$Mpc bins, allowing different choices for bin centres and widths.} in $s$, testing both 5 and 8 $h^{-1}$Mpc, and 0.01 in $|\mu|$. We then determine even moments of the redshift-space correlation function via 
\begin{equation}
\frac{2\xi_{\ell}(s)}{2\ell+1} = \sum^{100}_{i=1} 0.01\xi(s,\mu_i)L_{\ell}(\mu_i), 
\label{eq:xiell}
\end{equation}
where $\mu_i = 0.01i-0.005$ and $L_\ell$ is a Legendre polynomial of order $\ell$. In this work we only use the $\ell = 0$ moment. By defining the monopole this way, we ensure an equal weighting as a function of $\mu$ and thus a truly spherically averaged quantity. This means any distance scale we measure based on the BAO position in $\xi_0$ matches our definition of $D_V$ (given in Eq. \ref{DV_definition}).

\begin{figure}
\includegraphics[width=84mm]{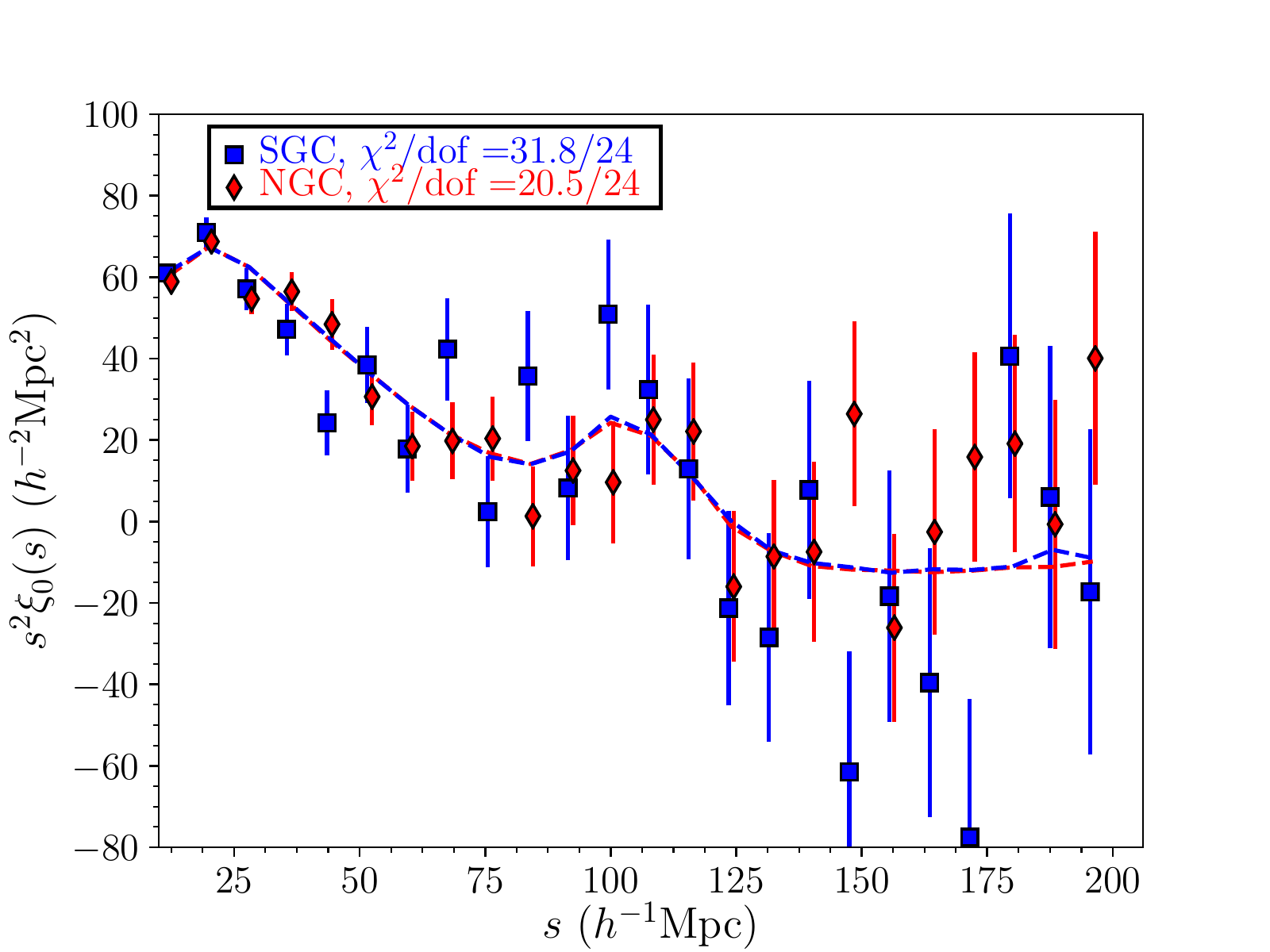}
\includegraphics[width=84mm]{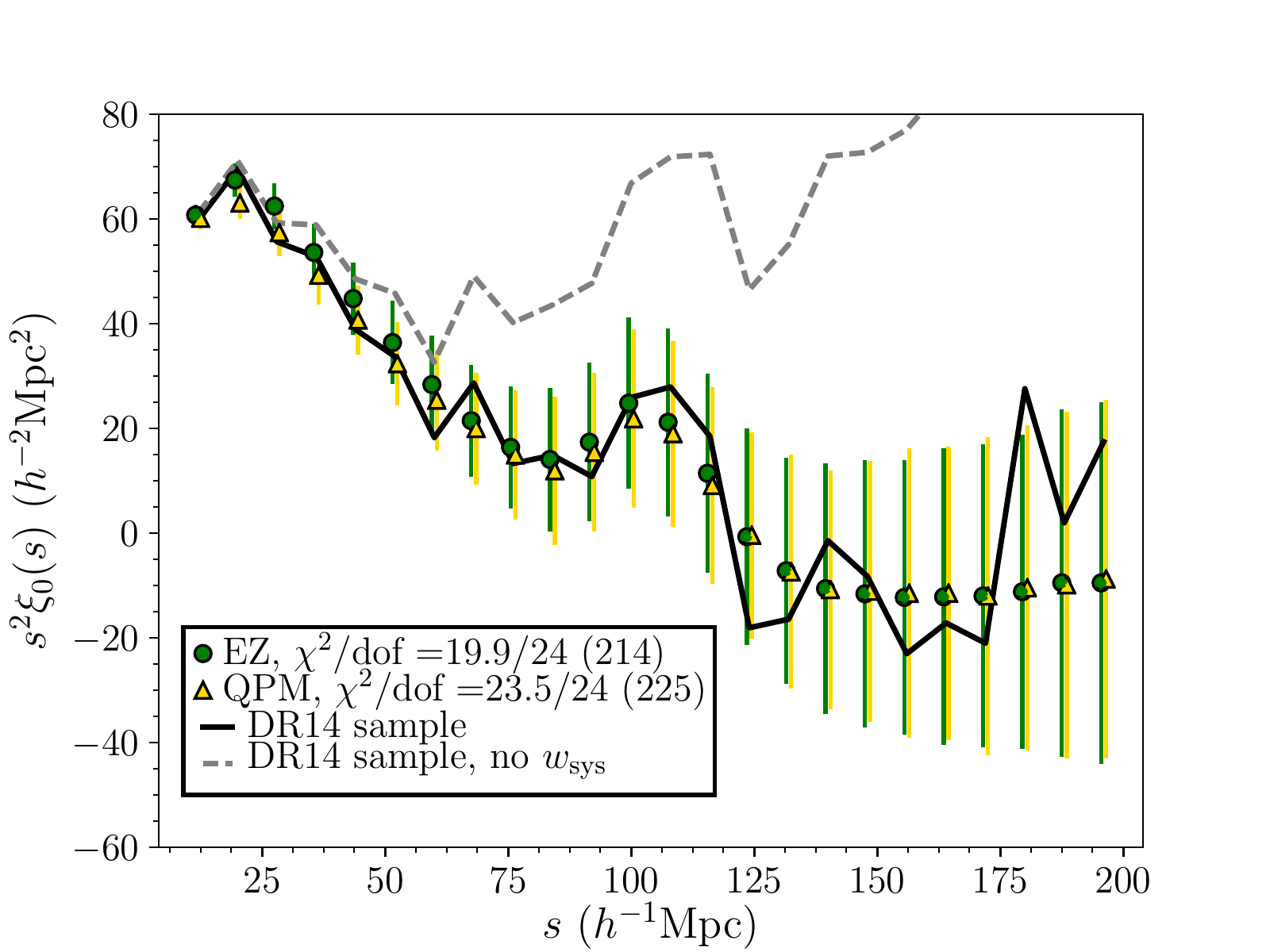}
  \caption{{\it Top panel}: The spherically averaged redshift-space correlation function of the DR14 quasar sample, for data in the SGC (blue squares) and NGC (red diamonds). The dashed curves display the mean of the 1000 EZmock samples. The data in each region are broadly consistent with the mean of the mocks and with each other. {\it Bottom panel}: The NGC and SGC data have been combined (solid black curve) and are now compared to both the EZ and QPM mocks (points with error-bars). The agreement is excellent. The dashed grey curve displays the result for the data when not applying systematic weights; the difference is dramatic and has $\chi^2$ significance of more than 180. The covariance matrix is dominated by the low number density of the DR14 quasar sample and the correlation between data points is low, e.g., the correlation between neighboring $s$ bins is $\sim$0.2. 
  }
  \label{fig:xiNScompEZ}
\end{figure}

The resulting correlation function is displayed in Fig. \ref{fig:xiNScompEZ}, where it is also compared to the mean of the mock samples we use. We describe the measurements further in Section \ref{sec:clustering}.

%subsection
\subsubsection{Fourier Space}

In order to measure the power spectrum of the quasar sample we start by assigning the objects from the data and random catalogues to a regular Cartesian grid. This is the starting point for using Fourier Transform (FT) based algorithms. In order to avoid spurious grid effects we use a convenient interpolation scheme to smooth the configuration-space overdensity field. 

We embed the entire survey volume into a cubic box with size $L_b =  7200\mpcoh$, and subdivide it into $N_g^3=1024^3$ cubic cells, whose resolution and Nyquist frequency are $7h^{-1}\,{\rm Mpc}$, and $k_{\rm Ny}=(2\pi/L_b) N_g/2=0.447\,h{\rm Mpc}^{-1}$, respectively. To obtain the smoothed overdensity field, an interpolation scheme is needed for the particle-to-grid assignment. By choosing a suitable interpolation scheme we can largely reduce the aliasing effect to a negligible level for frequencies smaller than the Nyqvist frequencies, which in this case comprises the typical scales for the BAO analysis.
Traditional interpolation schemes include the Nearest-Grid-Point (NGP), Cloud-in-Cell (CIC), Triangular-Shaped-Cloud (TSC) and Piecewise Cubic Spline (PCS). These options correspond to the zero-th, first, second and third order polynomial B-spline interpolations, respectively (see \citealt{Chaniotis04} for higher order interpolation schemes based on B-spline). Additionally, each of these interpolation schemes has an associated grid correction factor that has to be applied to the overdensity field in Fourier space \citep{Jing05}. 
The higher the order of the B-spline polynomial used in the grid interpolation, the smaller the effect of the grid on the final measurement. Aliasing arises as an extra limitation which cannot be avoided by just increasing the order of the grid interpolation scheme. Since for cosmological perturbations the bandwidth is not limited above a certain maximum cutoff frequency, the unresolved small scale modes are spuriously identified as modes supported by the grid, resulting in a contamination of the power spectrum, typically at scales close to the Nyqvist frequency.  Recently, \citet{interlacing} demonstrated that by displacing the position of the initial grid by fractions of the size of the grid cell the effect of the aliasing was greatly suppressed. This procedure is called {\it interlacing} and was originally presented in \citep{Hockney_Eastwood:1981}. In particular, \cite{interlacing} found that when a 2-step interlacing was combined with a PCS interpolation, the effect of aliasing was reduced to a level below $0.1\%$, even at the Nyquist scale. 

In this work, we apply a $5^{\rm th}$-order B-spline interpolation to calculate the overdensity field on the grid. Additionally, we combine two cartesian grids, displaced by half of their grid size, to account for the aliasing effect. We have checked (by doubling the number of grid cells per side) that the effect of aliasing is totally negligible in the range $k\lesssim0.4\,h{\rm Mpc}^{-1}$.

After applying the grid interpolation, we obtain an overdensity field $\Delta({\bf r}_i)$ at each grid centre, \citep{FKP}, 

\be\label{eq:Delta} \Delta({\bf r}_i)\equiv{w_{\rm tot}({\bf r}_i)}[n_{\rm qso}({\bf r}_i)-\gamma n_{\rm ran}({\bf r}_i)]/I^{1/2}_{2}.\ee
The quantity $w_{\rm tot}$ is the total weight for the quasars at the grid location given by Eq (\ref{eq:weight}), $n_{\rm qso}$ and $n_{\rm ran}$ are the number density at position $\rr$ of the quasars and random objects, respectively, $\gamma$ is the ratio between the total weighted numbers of the quasars ($N_{\rm qso}$) and random ($N_{\rm ran}$) catalogues, i.e., $\gamma = N_{\rm qso}/N_{\rm ran}$. Same as for the $\xi$ calculation, we use a random sample with 40$\times$ the size of the DR14 data set, 100$\times$ the size of the mean EZmock, and 70$\times$ the size of the mean QPM mock. Therefore, e.g., $\gamma\sim0.025$ for the data and  $\gamma\sim0.01$ for the EZmocks. The factor $I_2$, normalizes the amplitude of the observed power in accordance with its definition in a quasar distribution with no survey selection, 
\be
\label{eq:norm} I_2\equiv A \int  \langle w_{\rm sys} w_{\rm c} n_{\rm qso}\rangle^2(r)w_{\rm FKP}^2(r){\rm d}r
\ee
where, $\langle w_{\rm sys} w_{\rm c} n_{\rm qso}\rangle$ is the mean number density of quasars and $A$ the area of the survey in steradians. We perform this integration by sampling the mean number density of quasars in shells of $6.5 h^{-1}\,{\rm Mpc}$ and summing in the range $0.8\leq z\leq2.2$. 

In this work, we only present a measurement of the monopole (angle averaged with respect to the line of sight) of the power spectrum\footnote{Future eBOSS studies will use the anisotropic signal.}. To measure the power spectrum monopole, we must perform the Fourier transformation of the overdensity field $\Delta(\rr)$ defined in Eq (\ref{eq:Delta}). Since we are interested in the monopole, the varying line-of-sight of the quasars has no effect on our calculation.  Specifically, we need to calculate the following quantity, 
\be\label{eq:Flk} F_{0}(\k)\equiv\int d\rr \ \Delta(\rr)  e^{i \k\cdot\rr}.\ee
The power spectrum monopole is evaluated by a summation over $k$-directions and in the defined $k$-bin,
\begin{equation}
\label{eq:P0} P_0(k_{\rm eff})=\sum_i^{k-bin} F_{0}({\bf k}_i)F^{*}_{0}({\bf k}_i),
\end{equation}
where $k_{\rm eff}$ is the mean of all $|{\bf k}|$ values summed in the above equation.

\begin{figure}
\includegraphics[width=84mm]{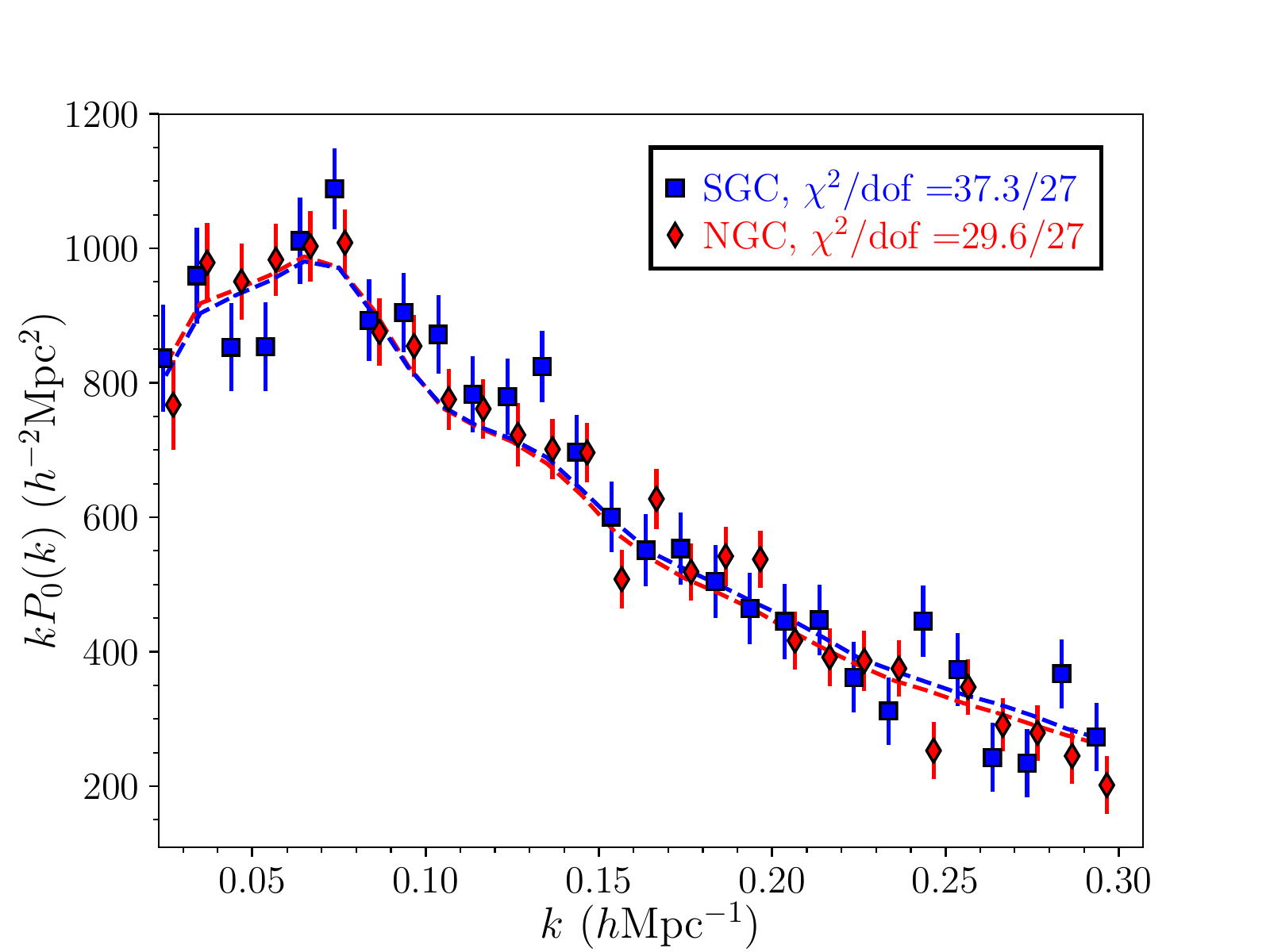}
\includegraphics[width=84mm]{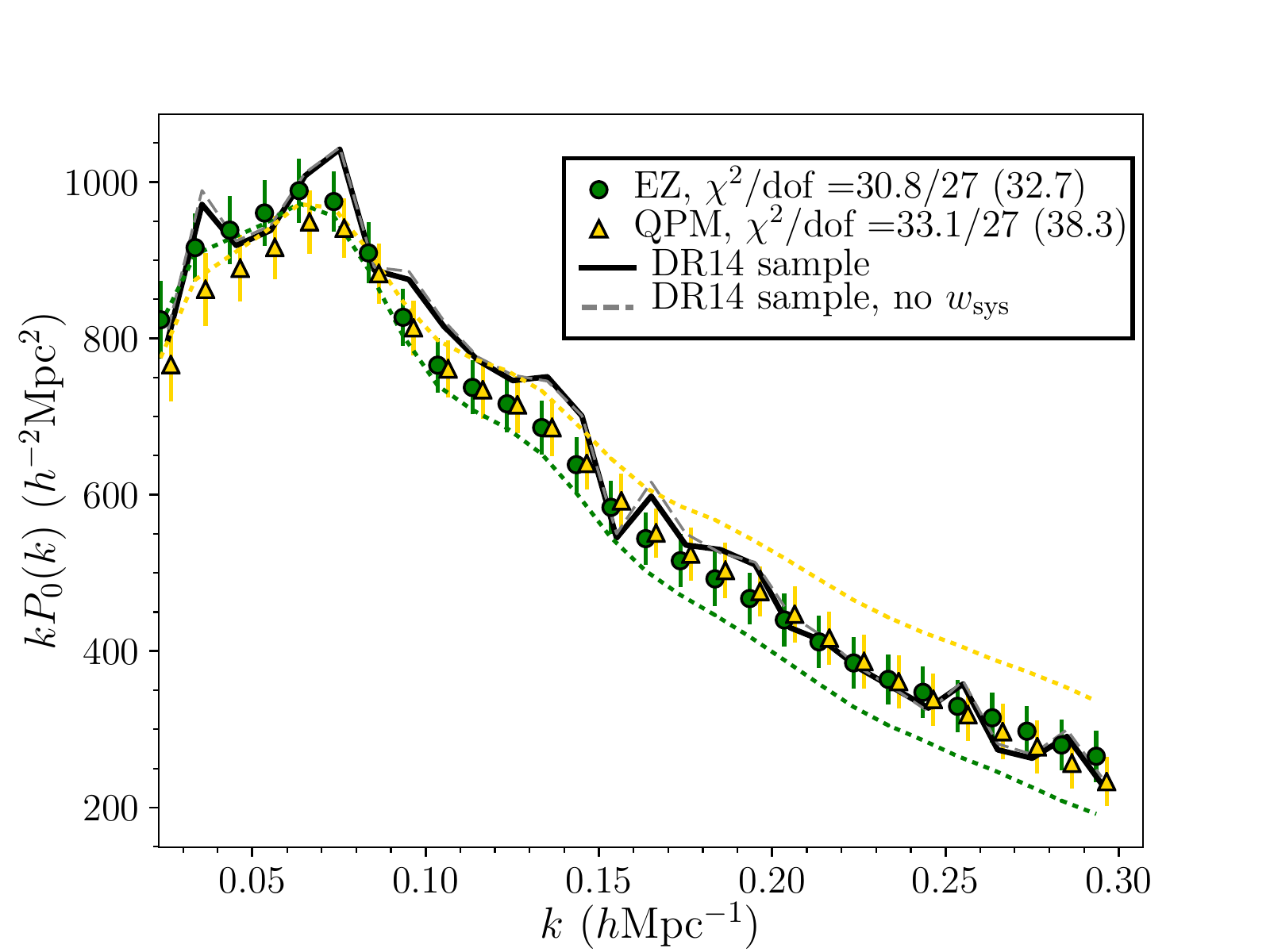}
  \caption{The same as Fig. \ref{fig:xiNScompEZ}, except for the power spectrum. For the points with error-bars, a constant of 350 has been subtracted from the mean of the QPM mocks and a constant of 250 has been added to the mean of the EZ mocks. The agreement would be poor without adding these constant offsets, as shown by the dotted curves, which display the respective means without them. In this case, there is clear disagreement at high $k$ with the DR14 measurements, but in opposite directions. A constant is marginalized over in the BAO analysis and we will explicitly demonstrate that our results are insensitive to the choice of mocks used for the covariance matrix, suggesting the constant offsets are unimportant.} We display results for $0.02 < k<0.30h$Mpc$^{-1}$, which is the range that will be used for any BAO measurements (with $k<0.23 $Mpc$^{-1}$ being our fiducial limit). 
  \label{fig:p0compmock}
\end{figure}

We perform the measurement of $P_0(k)$ binning $k$ linearly in bins of $0.01\,h~{\rm Mpc}^{-1}$ between $k=0$ and the Nyqvist Frequency. Within this wide range, we limit the BAO analysis to the frequencies $0.02 \leq k [h~{\rm Mpc}^{-1}] \leq 0.23$. Scales outside of this range contain negligible information on the BAO peak position. We have checked these statements by using the mock quasar catalogues. The resulting power spectrum contains 21 $k$-bins and is displayed in Fig. \ref{fig:p0compmock}, where it is also compared to the mean of the mock samples. We describe these results further in Section \ref{sec:clustering}.

%subsection
\subsection{BAO Modeling}

We use the same basic modeling template of the BAO signal for both configuration and Fourier space. The BAO model is determined in Fourier space and then either transformed to configuration space or passed through the window function in order to be compared to observations. For both approaches, we determine how different the BAO scale is in our clustering measurements compared to its location in a template constructed using our fiducial cosmology. There are two main effects with cosmological dependence\footnote{There is a third effect, which is a small shift in the BAO position due to non-linear evolution, described later in this section. It has minor dependence on cosmology and its total effect is negligible compared to the precision of our measurements, which we demonstrate in later sections.} that determine the difference between the observed BAO position and that in the template. The first effect is the difference between the BAO position in the true intrinsic primordial power spectrum and that in the model, with the multiplicative shift depending on the ratio $r_{\rm d}/r^{\rm fid}_{\rm d}$, where $r_{\rm d}$ is the sound horizon at the drag epoch (and thus represents the expected location of the BAO feature in co-moving distance units, due to the physics of the early Universe). The second effect is the difference in projection. The data are measured using a fiducial distance-redshift relation, matching that of the template: if  the actual cosmology is different than that assumed we expect a shift that depends on $H(z)$ in the radial direction, and $D_A(z)$ in the angular direction. For spherically averaged clustering measurements, we thus measure
\begin{equation}
\label{alpha_definition}\alpha =  \frac{D_V(z)r^{\rm fid}_{\rm d}}{D^{\rm fid}_V(z)r_{\rm d}},
\end{equation}
with
\begin{equation}
\label{DV_definition}D_V(z) = \left[cz(1+z)^2H(z)^{-1}D_A^2(z)\right]^{1/3}.
 \end{equation}
 Given sufficient signal-to-noise ratio, $D_A$ and $H(z)$ can be measured separately by using the isotropic and anisotropic signal. In this paper we only focus on the isotropic signal due to limited signal-to-noise ratio, and hence, we constrain $D_V$. 

The methodology we adopted to measure $\alpha$ is based on that used in \cite{alph} (and references therein). We generate a template BAO feature using the linear power spectrum, $P_{\rm lin}(k)$, obtained from {\sc Camb}\footnote{camb.info} \citep{camb,camb2} and a `no-wiggle' $P_{\rm nw}(k)$ obtained from the \cite{EH98} fitting formulae\footnote{In order to best-match the broadband shape of the linear power spectrum, we use $n_s = 0.963$, to be compared to 0.97 when generating the full linear power spectrum from {\sc camb}.} for $\xi$ and following \cite{Kirkby13} for $P(k)$, both using our fiducial cosmology (except where otherwise noted). 

Again emulating \cite{alph} (and references therein), given $P_{\rm lin}(k)$ and $P_{\rm nw}(k)$, the linear theory BAO signal is described by the oscillation pattern in the $\mathcal{O}_{\rm lin}(k)\equiv P_{\rm lin}(k)/P_{\rm nw}(k)$. We account for some non-linear evolution effects by `damping' this BAO signal:
\begin{equation}
\mathcal{O}_{\rm damp}(k) = 1+\left[\mathcal{O}_{\rm lin}(k)-1\right]e^{-\frac{1}{2}\Sigma_{\rm nl}^2k^2}.
\label{eq:odamp}
\end{equation}
This damping is treated slightly different in the $P(k)$ and $\xi(s)$ analyses, as we describe in Sections \ref{sec:ximod} and \ref{sec:Pkmod}. In addition to damping the BAO oscillations, non-linear evolution effects are also expected to cause small shifts (of order 0.5 per cent) in the BAO position \citep{PadWhite09}, which should have a small cosmological dependence (e.g., the size of the shift is likely dependent on $\sigma_8$). We will show that our results are insensitive to such effects.
 
\subsubsection{Correlation Function Modeling}
\label{sec:ximod}
For the correlation function, we simply use
\begin{equation}
P(k) = P_{\rm nw}(k)\mathcal{O}_{\rm damp}(k),
\end{equation}
and its Fourier transform in order to obtain the configuration-space BAO template, $\xi_{\rm temp}(s)$. We fix $\Sigma_{\rm nl} = 6h^{-1}$Mpc in the analysis and show that the results are insensitive to this choice. This choice is based on basic extensions of linear theory and the fact that the quasar sample has non-negligible redshift uncertainty. \cite{SeoEis07} provide predictions $\Sigma_{\perp} = 10.4D(z)\sigma_8$ and $\Sigma_{||} = (1+f)\Sigma_{\perp}$. This decomposition accounts for redshift-space distortions; the real-space prediction is given by $\Sigma_{\perp}$. The spherical average is $\Sigma^2_{\rm nl} = ([\Sigma^2_{\perp}]^2\Sigma^2_{||})^{1/3}$. For our fiducial cosmology, $\Sigma_{\perp} = 4.1h^{-1}$Mpc and $\Sigma_{\rm nl} = 5.2h^{-1}$Mpc. We compare these predictions to those obtained from real-space non-linear power spectrum predictions using \cite{Blas16}\footnote{\cite{Blas16} also includes the shift in the BAO peak, due to the non-linear growth of the matter power spectrum, which we evaluate in Section \ref{sec:mocktests}.}, evaluated at $z=1.5$ using {\sc FAST-PT} \citep{fastpt}. A value of $\Sigma_{\perp} =3.7h^{-1}$Mpc produces a BAO feature in our template defined by Eq. \ref{eq:odamp}, matching the amplitude of the \cite{Blas16} template, suggesting reasonable agreement with the more basic \cite{SeoEis07} approach. Thus, we expect $\Sigma_{\rm nl} \sim 5 h^{-1}$Mpc for redshift-space measurements; we increase this to $\Sigma_{\rm nl} = 6h^{-1}$Mpc in order to account for redshift uncertainties. We test and discuss this issue further in Section \ref{sec:mocktests}.

Given $\xi_{\rm temp}(s)$, we then fit to the data using the model
\begin{equation}
\xi_{0, {\rm mod}}(s) = B_0\xi_{\rm temp}(s\alpha) + A_{1} +A_2/s+A_3/s^2.
\label{eq:xi0mod}
\end{equation}
Including the polynomial makes our results insensitive to shifts in the broad-band shape of the measured $\xi_0$. As in previous analyses (e.g., \citealt{alph}), we apply a Gaussian prior of width 0.4 around the $B_0$ obtained when fitting $\xi_{\rm temp}$ to the data in the range $30 < s < 50h^{-1}$Mpc (not including the polynomial terms). These scales are safely outside of the scales where the BAO feature is significant. Using this prior ensures that the BAO feature in the model is neither unphysically large or small.

For both mocks and for the data, we adopt the appropriately weighted average of the NGC and SGC $\xi$ in order to obtain our BAO measurements. The configuration-space analysis does not have the Fourier-space window function concerns discussed in the following section. Thus, each correlation function BAO fit has five free parameters.

\subsubsection{Power Spectrum Modelling}
\label{sec:Pkmod}
In the power spectrum analysis, the position of the BAO peak is described by the oscillation pattern in $\mathcal{O}_{\rm lin}(k)$. The position of the peak is identified by shifting the pattern through the $\alpha$ parameter as $\mathcal{O}_{\rm lin}(k/\alpha)$.
We use the same power spectrum template form used for previous BAO fits in the BOSS survey \citep{Gil15BAO},
\begin{equation}
\label{Pbao_model}P(k,\alpha)=P_{\rm sm}(k)\left\{ 1+\left[\mathcal{O}_{\rm lin}(k/\alpha)-1\right]e^{-\frac{1}{2}\Sigma_{\rm nl}^2k^2}\right\}
\end{equation}
where the $P_{\rm sm}(k)\equiv B^2P_{\rm nw}(k)+A_1 k +A_2 + A_3/k$ accounts for all the non-linear and redshift space effects in the power spectrum monopole. It is possible to model $P_{\rm sm}(k)$ with higher polynomial coefficients such as $+A_4/k^2+A_5/k^3$. Although these terms were used for modelling the broadband power spectrum shape of the LRG galaxies at lower redshifts in BOSS \citep{Gil15BAO}, we have determined that for the current precision and redshift ranges  in this paper, adding these two extra terms does not affect the determination of $\alpha$ significantly.

The last step we need to incorporate in the model of Eq. \ref{Pbao_model} is the effect of the window function caused by the non-uniform angular distribution of quasars (see Fig. \ref{fig:footprint_QSO}), and the dependence of the mean density of quasars with the radial distance (see. Fig. \ref{fig:nz}). These two effects are accounted for by following the procedure described in \cite{Wilson_et_al:2016}.
The masked power spectrum, $\hat{P}_0$ is written as a Hankel Transform (HT) of the masked correlation function $\hat{\xi}_0$,
\begin{equation}
\hat{P}_0(k)=4\pi\int\hat{\xi}_0(s) j_0(sk)\, {\rm d}s,
\end{equation}
where $j_0$ is the spherical Bessel function, $j_0(x)=\sin(x)/x$, and $\hat{\xi}_0(s)$ can be written in terms of the correlation function $\ell$-multipoles, corresponding to the inverse HT of the un-masked power spectrum template model, 
\begin{equation}
\hat{\xi}_0(s)={\xi}_0(s) W_0^2+\frac{1}{5}{\xi}_2(s) W_2^2+\dots.
\end{equation}
We neglect any contribution of the power spectrum quadrupole into the monopole through the window function, and therefore we approximate, $\hat{\xi}_0(s)\simeq{\xi}_0(s) W_0^2$.
$W_i$ contains all the information on the radial and angular selection functions, and can be modelled either analytically or through the pair-counts of the random catalogue.
For simplicity we follow the later option and we write $W^2_0$ as,
\begin{equation}
\label{Weq}W_0^2(s)\propto \sum_{i,j}RR(s)/s^2,
\end{equation}
where $W_0^2(s)$ is normalised to 1 in the $s\rightarrow0$ limit. The $s^2$ term in the denominator accounts for the volume of the shell when the binning is linear in $s$.

\begin{figure}
\includegraphics[width=84mm]{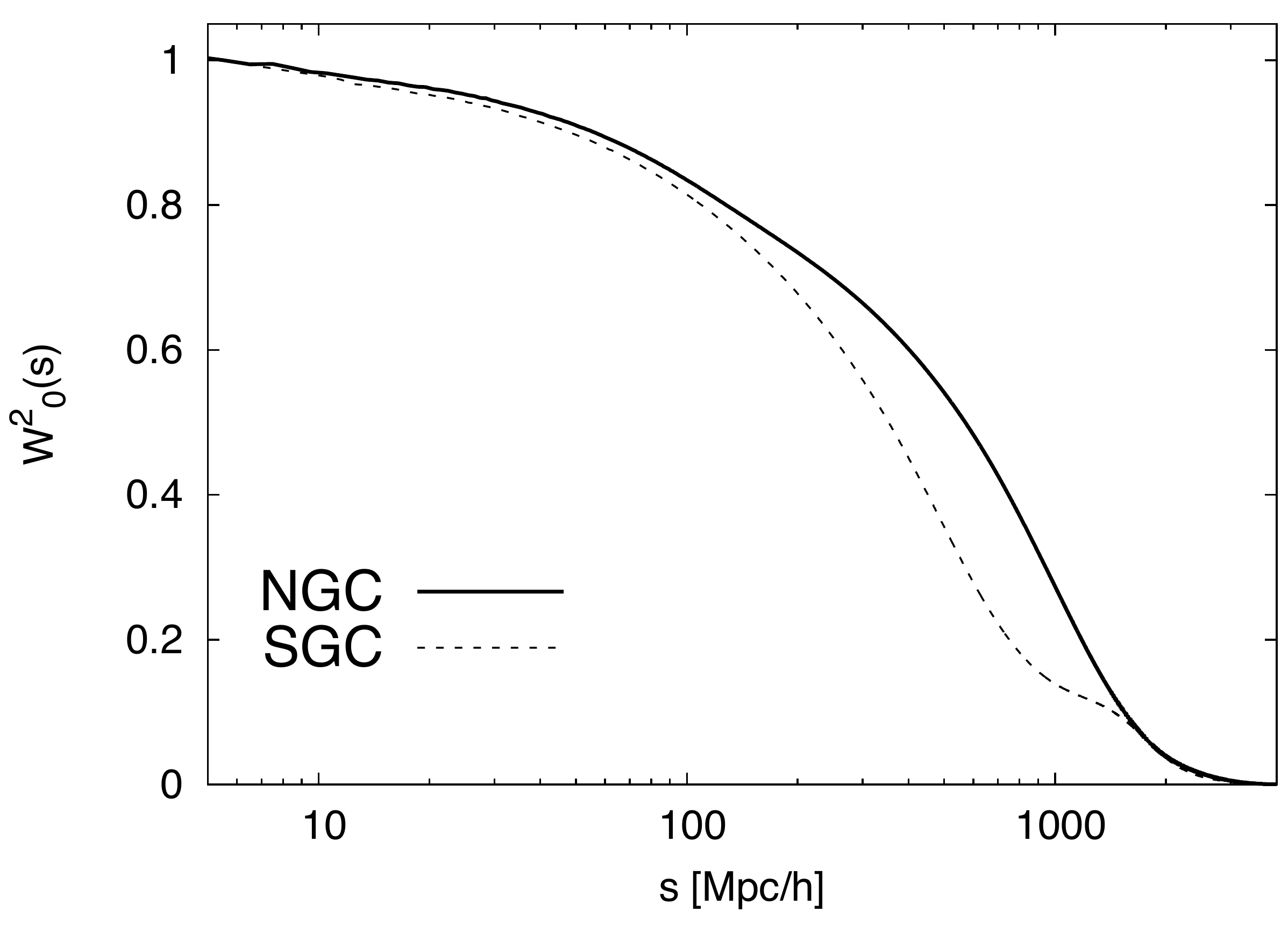}
\includegraphics[width=84mm]{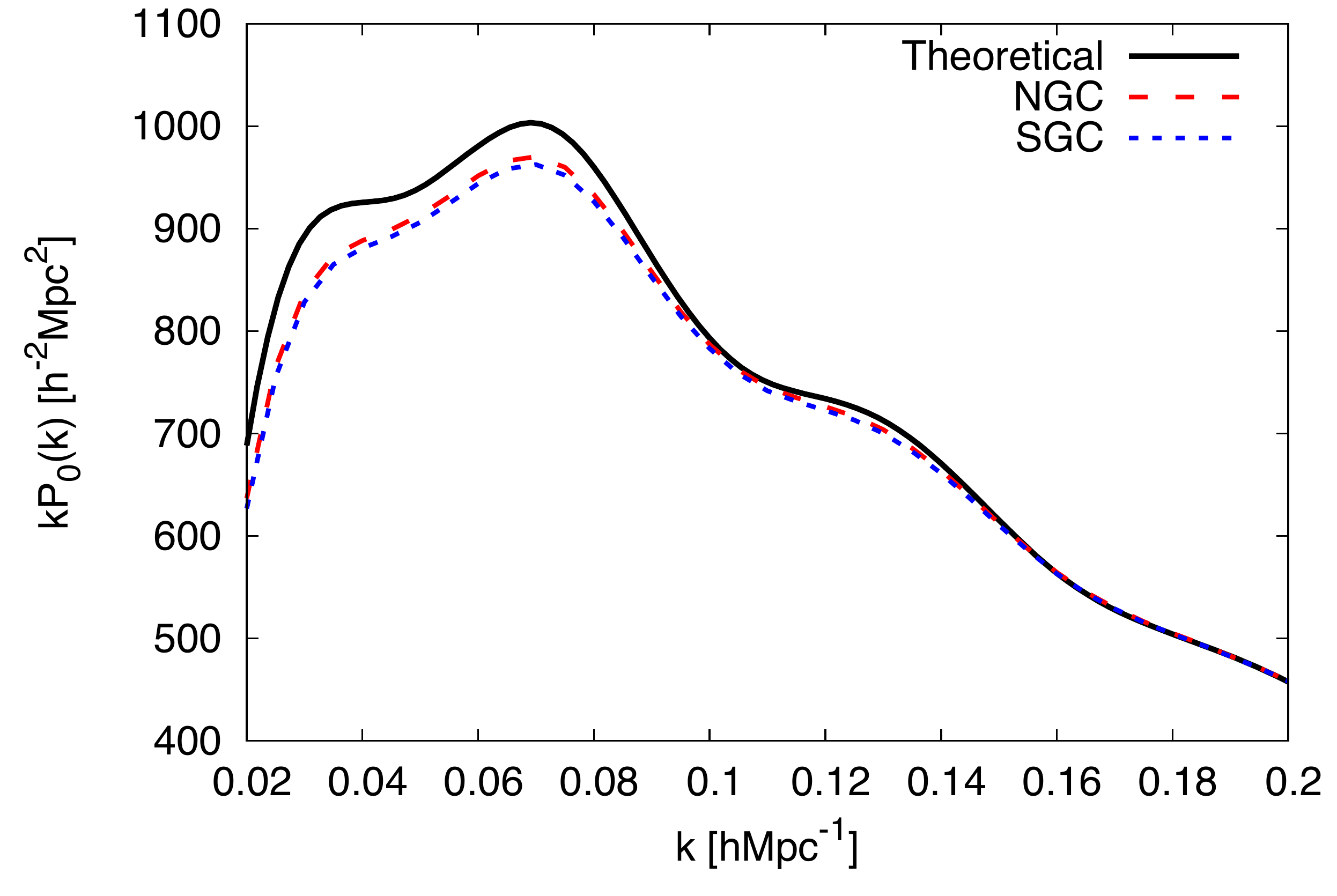}
  \caption{{\it Top panel}: The window function $W^2_0$ (see Eq. \ref{Weq}) for the NGC (solid line) and for the SGC (dashed line). {\it Bottom panel}: The convolution between the theoretical power spectrum with no survey effects (black solid line)  and the Fourier Transform of the window function $W^2_0$ (red dashed and blue dotted lines for the NGC and SGC geometries, respectively). The effect of the window is to damp the power at $k\lesssim0.15,h~{\rm Mpc}^{-1}$.}
  \label{fig:window}
\end{figure}

The top panel of Fig. \ref{fig:window} displays the performance of $W_0^2(s)$ for the redshift range $0.8\leq z \leq 2.2$, for the NGC and SGC patches, in solid and dashed black lines, respectively. In the case of no survey selection effect (for instance a periodic boundary condition simulation), the function $W_0^2(s)$ would approach $W_0^2(s)=1$, and the convolution between a theoretical power spectrum and the FT of $W_0^2(s)$ (a Dirac delta in this case) would produce no difference between the theoretical and the observed power spectrum. 
The effect of a non-flat redshift distribution and a non-uniform sky geometry produces a window function with the shape observed in the top panel of Fig. \ref{fig:window}. As expected, the departure from the ideal $W_0^2(s)=1$ case is more prominent in the SGC than in the NGS, as the SGC footprint covers a smaller angular area.  The effect of the window function is to produce an extra coupling term (in addition to the non-linear coupling) among the $k$-modes of the power spectrum. As a consequence, the covariance term is increased among $k$-modes and the amplitude of the observed power spectrum is reduced at large scales, as it is shown in the bottom panel of Fig. \ref{fig:window}. 

Since the shape of the window function is slightly different for the NGC and SGC regions, we choose to perform the power spectrum BAO fit separately, assuming no correlation among the two disconnected regions. Furthermore, we fit for different broadband parameters in the NGC and SGC, to account for different observational effects, such as photometric calibration, that can yield to a different effective biases for the two separate regions.  Thus, we fit for $B$, $A_1$, $A_2$ and $A_3$ separately for NGC and SGC, and keep the same value for $\alpha$ and $\Sigma_{\rm nl}$. Thus, in total we fit for 10 free parameters (9 if $\Sigma_{\rm nl}$ is kept constant) using 42 $k$-modes (21 for each patch) in the range $0.02\leq k\,[h{\rm Mpc}^{-1}] \leq 0.23$. We have observed that by following such approach the constraints on $\alpha$ (in both mocks and data) are improved (when compared to the case where the weighted average is used in the fit) and that the resulting likelihoods are closer to a Gaussian distribution.

%subsection
\subsection{Parameter Estimation}

We assume the likelihood distribution, ${\cal L}$, of any parameter (or vector of parameters), $p$, of interest is a multi-variate Gaussian:  
\begin{equation}
{\cal L}(p) \propto e^{-\chi^2(p)/2}.
\end{equation}
The $\chi^2$ is given by the standard definition
\begin{equation}
\chi^2 = {\bf D}{\sf C}^{-1}{\bf D}^{T},
\end{equation}
where ${\sf C}$ represents the covariance matrix of a data vector and ${\bf D}$ is the difference between the data and model vectors, when model parameter $p$ is used. We assume flat priors on all model parameters, unless otherwise noted.

In order to estimate covariance matrices, we use a large number of mock quasar samples (see Section \ref{sec:mocks}), unless otherwise noted. The noise from the finite number of mock realizations requires some corrections to the $\chi^2$ values, the width of the likelihood distribution, and the standard deviation of any parameter determined from the same set of mocks used to define the covariance matrix. These factors are defined in \cite{Hartlap07,Dod13} and \cite{Per14}; we apply the factors in the same way as in, e.g., \cite{alph}. For our fiducial $\xi(s)$ results, we use 1000 mocks and 18 measurement bins (fitting to the weighted mean of the NGC and SGC results). For the $P(k)$ analysis we fit NGC and SGC separately, which corresponds of using 1000 mocks and 21 measurement bins for each NGC and SGC regions. In both cases, the number of mock realisations is much larger than the number of measurement bins, implying the finite number of mocks has less than a 2 per cent effect on our uncertainty estimates. We observe that for both $\xi(s)$ and $P(k)$ analyses the corresponding covariance matrices are dominated by their diagonal elements.

 The covariance matrix derived from the mocks is dominated by its shot noise component due to the low density of the quasars. On one hand, the EZ and QPM mocks have been produced to match the effective number of objects observed in the data sample (${\bar{N}}_{\rm eff}$ in Table \ref{tab:catalogues}). On the other hand, the EZ and QPM mocks do not currently include the redshift failures and collision pair effects. Therefore, the actual number of quasars in the mocks matches the effective number of quasars on the data catalogue, which makes the total number of quasars in the mocks slightly higher than in the data. As a consequence, the diagonal and off-diagonal components of the covariance are underestimated approximately by the ratio between ${\bar{N}}_{\rm eff}$ and ${\bar{N}}_{\rm Q}$ in the range $0.8\leq z \leq 2.2$. In order to correct for this effect we re-scale the derived covariance elements by these factors when analyzing the DR14 data, which are 1.069 for the NGC and 1.073 for the SGC. Future eBOSS analyses will include these effects in the mocks (and therefore this correction will be unnecessary). Here, we use the simple scaling as it has only a 3 per cent effect on the recovered uncertainty.

%section
\section{Simulated Catalogs}
\label{sec:mocks}

We use two different methods to create a total of 1400 simulations of the DR14 quasar sample, which we refer to as `mocks'. In order to create this number of mocks, approximate methods are required. Our approach in this respect is similar to previous BOSS analyses \citep{Manera13,alph,Acacia}. The two methods used are `EZmock' and `QPM' and are described in the following sub-sections.

\subsection{EZmocks}
\label{sec:ezmock}
For this work, we construct 1000 light-cone mock catalogues covering the full survey area of DR14 (NGC+SGC) and reproducing the redshift evolution of the observed quasar clustering. These are created using the `EZmock' (Effective Zel'dovich approximation mock catalogue, \citealt{EZmock}) method. EZmocks are constructed using the Zel'dovich approximation of the density field. This approach accounts for non-linear effects and also halo bias (i.e. linear, nonlinear, deterministic, and stochastic bias) into an effective modeling with few parameters, which can be efficiently calibrated with observations or N-body simulations. \cite{Chuang15mockcomp} demonstrates that the EZmock technique is able to precisely reproduce the clustering of a given sample (including 2- and 3-point statistics) with minimal computational resources, compared to other methods. We use an improved version of EZmock code with respect to the one described in \cite{EZmock}. In our work, we assign the positions of quasars to simulated dark matter particles instead of populating them following a cloud-in-cell distribution. With this change, we do not need to enhance the BAO signal in the initial conditions, as done in \cite{EZmock}.

For this study, we calibrate the bias parameters with the observed DR14 eBOSS quasar clustering directly. The NGC and SGC regions are created from separate simulations and are treated independently, with bias values fit to the measured clustering and the $n(z)$ taken as in Fig. \ref{fig:nz}. Comparisons between the mean clustering in the EZmock samples and the measured eBOSS clustering can be found in Figs. \ref{fig:xiNScompEZ} and \ref{fig:p0compmock}, demonstrating that a good match has been produced. The EZmocks use the same initial power spectrum used by the mock catalogues of the final BOSS data release (DR12; \citealt{Kitaura15,Acacia}). The fiducial cosmology model is $\Lambda$CDM with $\Omega_{m}$=0.307115, $h$=0.6777, $\sigma_8$=0.8225, $\Omega_{b}$=0.048206, $n_s$=0.9611 (see \citealt{Kitaura15} for details). 

Each light-cone mock constructed for this work is composed of 7 redshift shells. The redshift shells for a given light-cone mock are computed using different EZmock parameters but they share the same initial Gaussian density field so that the background density field is continuous. Each redshift shell is taken from one corresponding EZmock periodic box with the size of (5$h^{-1}$Gpc)$^3$. For this study, we generated 1000$\times$7(shells)$\times$2(NGC+SGC)$=$14,000 EZmock boxes in total. We use the code {\it
make\_survey}~\cite{carlson16,QPM} to construct each redshift shell from the corresponding box.

\begin{figure}
\centering
\includegraphics[width=1. \columnwidth,clip,angle=-0]{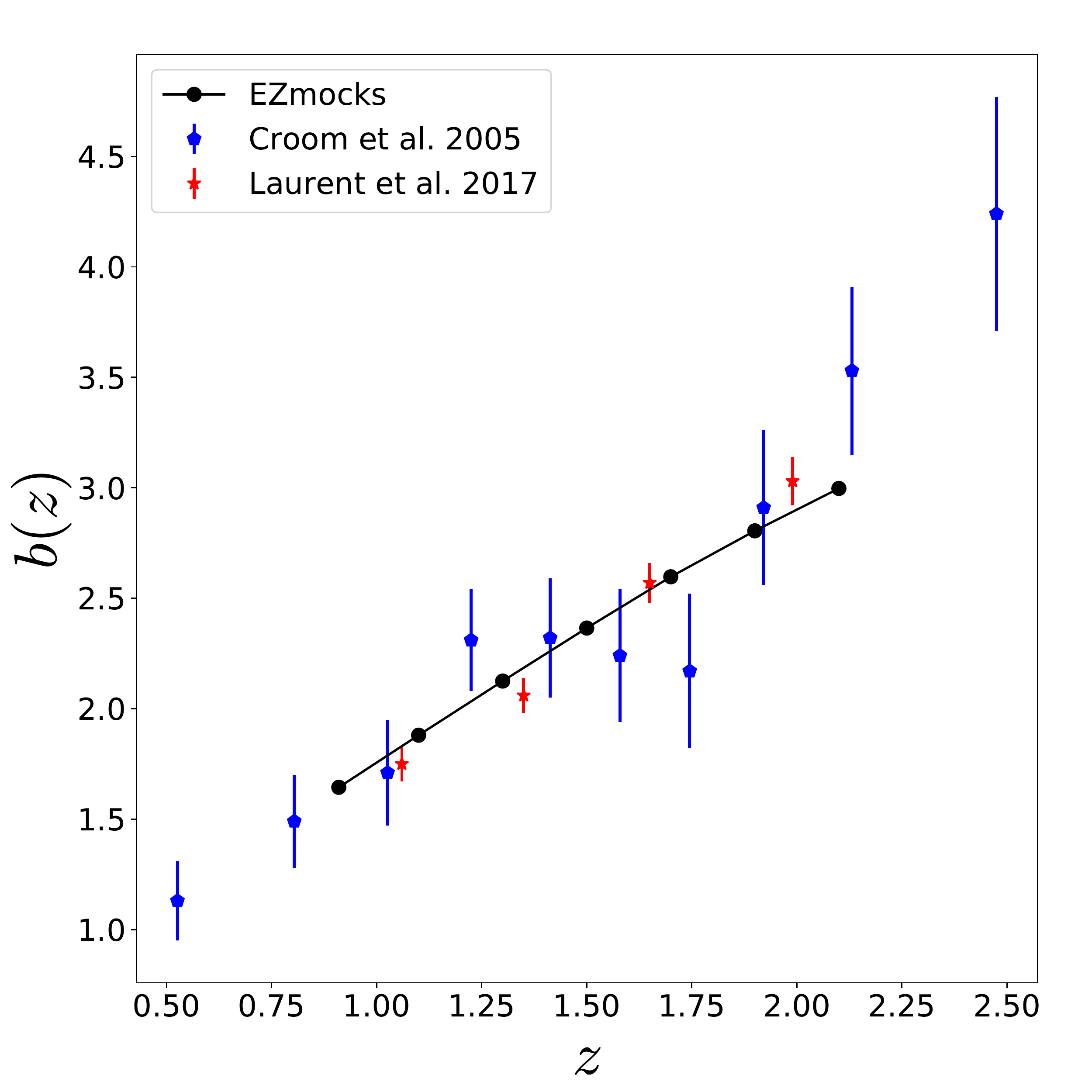}
\caption{
We compare the linear bias measured from EZmocks used in this work with the other measurements from observed data \citep{Croom05,Laurent16}. The bias evolution in the EZmocks is in good agreement with these works. 
}
\label{fig:EZmock_bz}
\end{figure}

In order to determine the redshift evolution of each EZmock parameter, we need to measure/determine the parameters at different redshifts. However, the clustering measurements from the observation corresponding to each redshift shell are too noisy to be used to determine the EZmock parameter values. Therefore, we use samples from overlapping redshift bins to measure the EZmock parameters at different redshift and do a proper inter- or extra- polation to determine the parameters for all the 7 redshift shells.
For each EZmock parameter $p$, we assume that its functional dependency is well approximated by
\begin{align}
p=c_0 + c_1 Z(1) + c_2 Z(2), \label{eq:EZspline}
\end{align}
where
\begin{align}
Z(i)=\frac{\sum\limits_{z_{min}\leq z_j < z_{max}} N(z_j) {z_j}^i}{\sum\limits_{z_{min} \leq z_j < z_{max}} N(z_j)},\quad i=1,2,
\end{align}
of the number count $N(z_j)$ within a given redshift bin $z_j$ (with bin size $\Delta z=0.01$).

For this work, we measured the EZmock parameters from observation in the following three redshift ranges: $0.8 \leq z_1 < 1.5$, $1.2 \leq z_{2} < 1.8$, and $1.5 \leq z_{3} < 2.2$, respectively. We determine the $c_0$, $c_1$, and $c_2$ by solving the following equations

\begin{align}
p_{1}=c_0 + c_1 Z_{1}(1)+ c_2 Z_{1}(2),\nonumber\\
p_{2}=c_0 + c_1 Z_{2}(1) + c_2 Z_{2}(2),\nonumber\\
p_{3}=c_0 + c_1 Z_{3}(1) + c_2 Z_{3}(2). \label{eq:EZsplineSys}
\end{align}

Having solved the system of Eqs. \ref{eq:EZsplineSys} for the coefficients $c_0$, $c_1$ and $c_2$, Eq. \ref{eq:EZspline} is used to determine an EZmock parameter value for any given redshift shell.

Fig. \ref{fig:EZmock_bz} compares the linear bias measured from EZmocks used in this work with the other measurements  \citep{Croom05,Laurent16}. For each redshift shell, the bias is measured from the 1000 EZmock boxes. The EZmock parameters have been calibrated using the observed data in three redshift bins as described above, so that EZmocks present a consistent linear bias evolution in comparison with other works.
More details of the algorithm, clustering performance, and tests on bias evolution will be provided in \cite{Chuang17}.

\subsection{QPM Mocks}
\label{sec:QPM}
\begin{figure*}
\includegraphics[width=164mm]{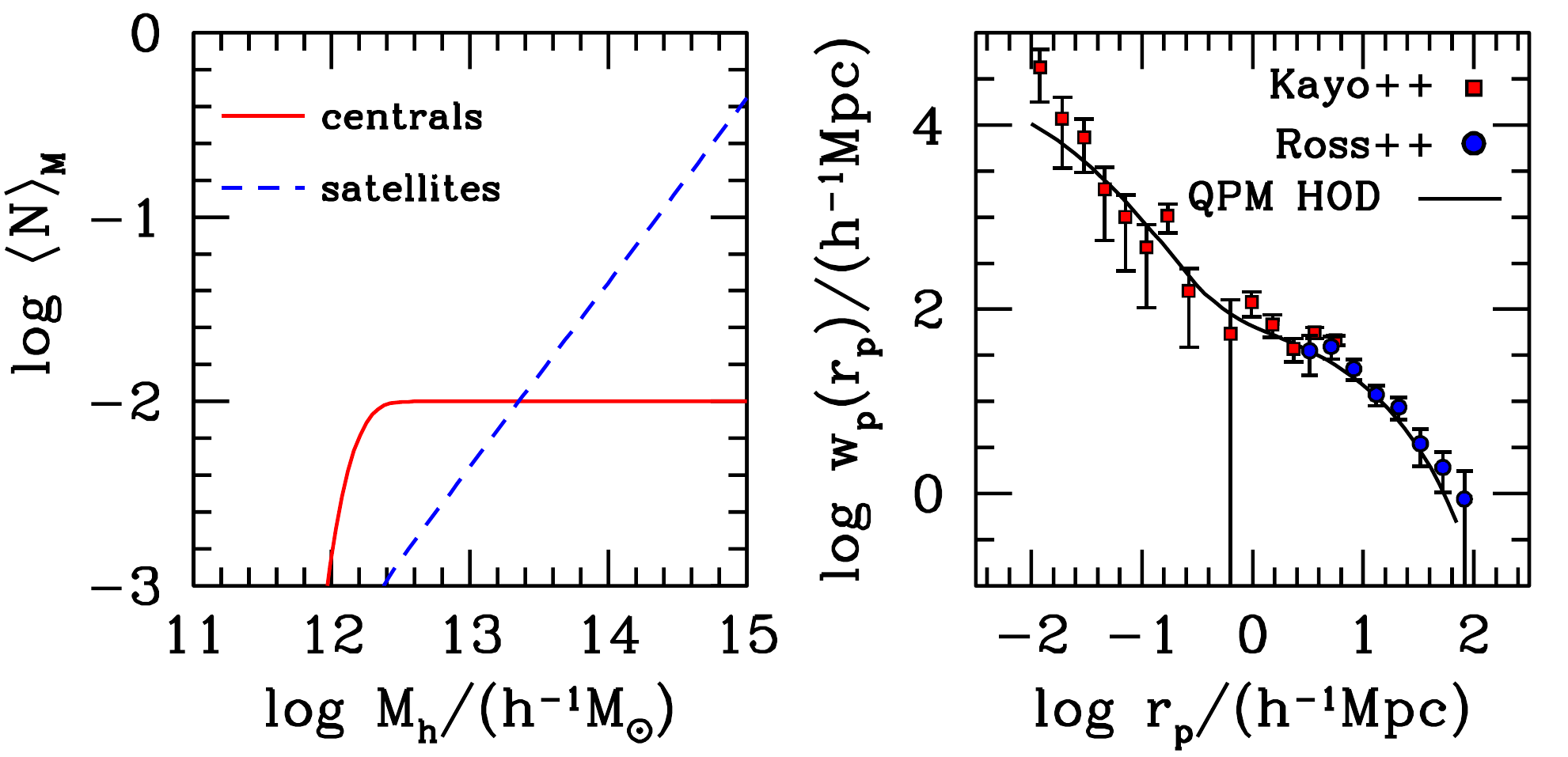}
\caption{{\it Left panel:} The mean halo occupation of quasars used in the QPM mocks. The solid red curve shows the mean number of central quasars per halo, while the blue dashed curve indicates the mean number of satellite quasars per halo. The horizontal asymptote of 0.01 central quasars per halo reflects the adopted duty cycle of 1\% for quasars in the eBOSS sample. The number of central and satellite quasars in a halo are uncorrelated with one another; i.e., the presence of a satellite quasar is not conditioned on the existence of a central quasar. {\it Right panel:} The projected correlation function, $w_p(r_p)$, yielded by the HOD in the left panel. For comparison, the red squares are the measurement of $w_p(r_p)$ of quasars from \citet{kayo:12}, while the blue circles represent the measurements from \citet{ross_etal:09}. The solid curve is calculated from the theoretical HOD model of \citet{tinker_etal:12}.
%For comparison, red squares \citep{kayo:12} and blue circles \citep{ross_etal:09} display measurements of $w_p(r_p)$ of quasars. The solid curve is calculated from the theoretical HOD model of \citet{tinker_etal12}.
%For comparison, the red squares are the measurement of $w_p(r_p)$ of quasars from \cite{kayo:12}, while the blue circles represent the measurements from \cite{ross_etal:09}. The solid curve is calculated from the theoretical HOD model of \cite{tinker_etal12}.
}
 \label{fig:qpm}
\end{figure*}

Quick Particle Mesh (QPM) mocks follow the method of \cite{QPM}. In brief, QPM uses a low-resolution particle mesh gravity solver to evolve a density field forward in time. This approach captures the non-linear evolution of the density field, but does not have sufficient spatial or mass resolution to form dark matter halos. Particles are sampled from the density field in a fashion to approximate the distribution of small-scale densities of halos. This process mimics both the one-point and two-point distributions of halos---their mass function and bias function. This approximate halo catalog can then be treated in the same manner as that from a high-resolution simulation and populated with galaxies according to a halo occupation distribution (HOD; c.f. \citealt{tinker_etal:12}). For the mocks used in this paper, we have adjusted the parameters of \cite{QPM} that map local density to halo mass to account for the change in redshift ($z\sim 2.5$ rather than $z\sim 0.5$ in \citealt{QPM}), as well as extending this mapping to lower halo masses, as required by the halos occupied by quasars.

To parameterize the halo occupation of quasars, we use the five-parameter HOD presented in \cite{tinker_etal:12}, which separates objects into central quasars and satellite quasars. To determine the HOD of the DR14 quasar sample, the peak of the $n(z)$ curve is the number density that the HOD is required to match. This number, in addition to the quasar large-scale bias measurement of $b_{\rm Q}=2.45$ allows us to determine the duty cycle of quasars. In the HOD context, the duty cycle is the fraction of halos that have quasars at their centres. There is some extra freedom due to the fraction of quasars that are satellites in larger halos, but this population is a minority of quasars. The left-hand side of Fig. \ref{fig:qpm} shows the HOD that matches both the peak of $n(z)$ and $b_{\rm Q}=2.45$ \citep{Laurent16} with a satellite fraction of 0.15. The right-hand side of Figure \ref{fig:qpm} compares the projected correlation function predicted by this HOD to measurements of quasar clustering from \cite{kayo:12} and \cite{ross_etal:09}. Although the quasar selection algorithms for these two papers are markedly different from eBOSS DR14, both yielding much smaller number densities than the DR14 sample, the measured clustering is remarkably similar to that predicted by the HOD model. The smaller number density in the \cite{ross_etal:09} and \cite{kayo:12} samples would be reflected in a lower duty cycle, with limited impact on the amplitude and shape of clustering. 

The implementation of the HOD used here assumes no correlation between the occupation of central and satellite quasars; i.e., the presence of a satellite quasar is not conditioned on the existence of a central quasar in a given halo. This assumption is also borne out by Fig. \ref{fig:qpm}---if all satellites were required to exist in halos with a central quasar, the amplitude of $w_p(r_p)$ at $r_p\la 1 h^{-1}$Mpc would be larger by a factor of 100 while the large-scale bias would be unchanged. Such a dramatic change in the shape of quasar clustering is clearly not allowed by the data. 

We simulated 100 cubic boxes of size $L=5120h^{-1}\,{\rm Mpc}$. The boxes are remapped to fit the volume of the full planned survey using the code {\sc make\_survey}~\cite{carlson16,QPM}. The same 100 cubic boxes are used for the Northern and Southern Galactic caps. The analysis presented here is based on the first two years of eBOSS data and therefore on a smaller volume. This situation allows to use different parts of a single QPM box to produce different realizations. Changing the orientation of the original box, we identified four configurations with less than 1.5 per cent overlap. We used these configurations to produce 400 QPM mocks per Galactic cap. The overlap between SGC and NGC can be as high as 10 per cent, but we identified pairs of configurations where the overlap is kept below 2 per cent. In this way we are able to compare Fourier space and configuration space results for the whole survey on a mock-by-mock basis (see Section~\ref{sec:mocktests}).

We transform the boxes to mock catalogs in the following manner. Cartesian comoving coordinates of the cubic boxes are transformed to angular coordinates and radial distances using a flat $\Lambda$CDM cosmology defined in Table~\ref{tab:cosmo}. Objects outside the angular mask are removed and veto masks accounting for bright stars, bright objects, plates centerposts and bad photometric fields are applied in the same manner as for the data (see Section \ref{sec:veto_masks}). The number density of quasars is downsampled to fit the redshift distribution (see Fig. \ref{fig:nz}) and FKP weights are calculated using the same value of $P_0$ (see Eq.~\ref{eq:wfkp}). Furthermore, redshifts are smeared according to a Gaussian distribution of width taken from the eBOSS early analysis~\cite{eboss}, namely $\sigma_z=300\,$km s$^{-1}$ for $z<1.5$ and $\sigma_z=400\times(z-1.5)+300\,$km s$^{-1}$ above.

%section
\section{Tests on Mocks}
\label{sec:mocktests}
We test our methodological choices by analyzing our mock catalogs and the robustness of the results to these different choices. These tests inform our decisions about how to combine results from different clustering estimators. When quoting uncertainties, we use half the width of the $\Delta \chi^2 = 1$ region, matching the approach of \cite{Ross15}. This choice best extrapolates to the expectation for data with greater signal to noise (e.g., future data releases), as the likelihood for BAO measurements is generally wider than a Gaussian distribution\footnote{This effect is negligible for high signal-to-noise ratio measurements, such as those in \cite{Acacia}.}. 

We first consider the results obtained from the mean of the mock samples, which are listed in Table \ref{tab:baomockmean}. For $\alpha$, we subtract the expected value of 1.0010 from the EZ mocks and and 1.0011 for the QPM mocks. All of the mean mock results are biased to be high, by between 0.001 and 0.004. The correlation function and power spectrum results show good agreement, with negligible differences in $\alpha$ and its uncertainty for our fiducial cases. 

Non-linear structure growth is expected to produce such slightly biased $\alpha$ values, of expected fractional size $\sim 0.5D(z)^2$ \citep{PadWhite09}, implying a shift of 0.13 per cent for the DR14 quasar sample. We test this assumption by using \cite{Blas16} to produce a BAO template that incorporates the expected non-linear structure growth effects. We compare this result to our template given by Eq. \ref{eq:odamp} with $\Sigma_{\rm nl} = 3.7h^{-1}$Mpc, as such a value yields a BAO feature of matching size. We measure a shift between the two templates of 0.07 per cent, which represents the prediction for the shift in $\alpha$ for the matter power spectrum. \cite{PadWhite09} demonstrate the expected shift is roughly proportional to the bias of the sample, as the shift can be explained by higher-order bias terms and these for dark matter halos are roughly proportional to the linear bias. Given that \cite{Laurent16} obtains a bias of 2.45 for eBOSS quasars, we should expect roughly 0.2 per cent shifts in the $\alpha$ recovered from our fiducial template (which includes no shifts due to non-linear structure growth). This prediction is generally consistent with our results when using the mean of the mocks. The differences in these tests, compared with the expected $\alpha$, are at most 2$\sigma$ in terms of the ensemble of mocks being tested and never more than 0.1$\sigma$ compared to the expected result for a single realization. We thus consider any potential bias due to non-linear structure growth to be negligible given the uncertainty obtained from the DR14 quasar sample.

In order to test the effect of redshift uncertainties, we have run the QPM mocks with a variety of assumptions on the redshift uncertainty; full details can be found in \cite{Zarrouk17}. We simply consider the fiducial case, with redshift uncertainty matching Fig. 11 of \cite{eboss} and a case without any redshift uncertainty. Without any redshift uncertainty, the mean $\xi(s)$ of the QPM mocks is best-fit by $\Sigma_{\rm nl} = 3.7h^{-1}$Mpc; with the redshift uncertainty, the best-fit is $\Sigma_{\rm nl} = 5.5h^{-1}$Mpc. These results suggest our choice to use $\Sigma_{\rm nl} = 6.0h^{-1}$Mpc is slightly conservative (greater $\Sigma_{\rm nl}$ leads to a greater uncertainty obtained from the likelihood).

\begin{table}
\centering
\caption{Tests of BAO fits on the mean of the quasar mocks. Fiducial results use the EZmocks for the covariance matrix, $\Delta s = 8h^{-1}$Mpc; $35 < s < 180h^{-1}$Mpc for $\xi(s)$, and evenly-spaced $\Delta k = 0.01h$Mpc$^{-1}$; $0.02 < k < 0.23\,h$Mpc$^{-1}$ for $P(k)$. The uncertainties are those for the full ensemble, e.g., one with 1000 (400) times the size of a single EZ (QPM) mock. The EZmock and QPM results are independent, but results using the same set of mocks are highly correlated. A small positive bias is expected from non-linear structure growth, see text for details.}
\begin{tabular}{lc}
\hline
\hline
case  & $\alpha-\alpha_{\rm exp}$\\
\hline
EZ mocks:\\
~~~~~$\xi(s)$:\\
fiducial & $0.0023\pm0.0016$\\
 5$h^{-1}$Mpc & $0.0027\pm0.0016$ \\
 ~~~~~ $P(k)$:\\
fiducial & $0.0019\pm0.0017$  \\
$k_{\rm max}=0.30\,h{\rm Mpc}^{-1}$ & $0.0009\pm0.0017$  \\
$\Sigma_{\rm nl} =  [6\pm3]\,h^{-1}{\rm Mpc}$ & $0.0021\pm0.0016$  \\
$\Sigma_{\rm nl} = [6\pm3]\,h^{-1}{\rm Mpc}$ \&  $k_{\rm max}=0.30$  & $0.0011\pm0.0016$  \\
log$k$ - binning & $0.0032\pm0.0017$  \\
log$k$ - binning \& $k_{\rm max}=0.30\,h{\rm Mpc}^{-1}$  & $0.0022\pm0.0016$  \\
$A_4$, $A_5$ terms  & $0.0037\pm0.0017$  \\
\hline
QPM mocks:\\
~~~~~$\xi(s)$:\\
fiducial & $0.0017\pm0.0028$ \\
    5$h^{-1}$Mpc& $0.0027\pm0.0028$ \\
QPM cov & $0.0023\pm0.0026$ \\
~~~~~$P(k)$:\\
fiducial  & $0.0017\pm0.0027$  \\
QPM cov  & $0.0012\pm0.0026$  \\
\hline
\label{tab:baomockmean}
\end{tabular}
\end{table}

Table \ref{tab:baomock} displays the results obtained when fitting each individual mock. Here, we only report statistics for mocks that have a $\Delta \chi^2 =1$ on both sides of the minimum $\chi^2$, within our prior range of $0.8 < \alpha < 1.2$. We designate each such mock result a `detection'. Over 90 per cent of the mock samples satisfy this condition. For this reason, the mean $\sigma$s are slightly less than that obtained for the mean of the mock realizations (multiplying the uncertainty in Table \ref{tab:baomockmean} by $\sqrt{N_{\rm mock}}$). The standard deviation of these samples is close to the mean uncertainty, as would be expected for a Gaussian distribution. 

\begin{table}
\centering
\caption{Statistics for BAO fits on EZ and QPM mocks. $\langle\alpha\rangle$ is the mean measured BAO parameter  with $1\sigma$ bounds within the range $0.8 < \alpha < 1.2$. $\langle\sigma\rangle$ is the same for the uncertainty obtained from $\Delta\chi^2=1$ region and $S$ is the standard deviation of these $\alpha$. $N_{\rm det}$ is the number of realizations with such $1\sigma$ bounds. Both EZmocks and QPMmocks were created with a slightly different cosmology than our fiducial assumed cosmology. Thus, for the EZmocks, the expected $\alpha$ values are 1.0010; and for QPM 1.0011. The $\xi$ bin size is 8$h^{-1}$Mpc, unless noted otherwise. Tests of shifting bin centres are noted by $+x$, with $x$ representing the shift in $h^{-1}$Mpc. The damping is $\Sigma_{\rm nl} = 6h^{-1}$Mpc, unless otherwise noted. The binning for $P(k)$ is linear spaced in 0.01$h$Mpc$^{-1}$, unless otherwise noted. As for $\xi$, test of shifting bin centres are noted by $+x/4$, with $x/4$ representing the shift in fractions of the $k$-bin size. The EZmock and QPM results are independent, but results using the same set of mocks are highly correlated. A small positive bias is expected from non-linear structure growth, see text for details.}
\begin{tabular}{lccccc}
\hline
\hline
case (+bin shift)  & $\langle\alpha\rangle$ & $\langle\sigma\rangle$ & $S$ & $N_{\rm det}/N_{\rm tot}$\\
\hline
EZ mocks:\\
{\bf consensus} $P(k)+\xi(s)$ & 1.003 & 0.050 & 0.050 & 944/1000\\
~~~~~$\xi(s)$:\\
combined & 1.003 & 0.049 & 0.049 & 939/1000\\
fiducial & 1.002 & 0.048 & 0.050 & 932/1000\\
 +2  & 1.002 & 0.049 & 0.050 & 928/1000\\
 +4  & 1.002 & 0.048 & 0.050 & 938/1000\\
 +6  & 1.003 & 0.048 & 0.051 & 929/1000\\
 5$h^{-1}$Mpc & 1.003 & 0.049 & 0.050 & 937/1000\\
~~~~~ $P(k)$:\\
combined & 1.002 & 0.052 & 0.050 & 941/1000\\
fiducial & 1.002 & 0.051 & 0.051 & 929/1000 \\
+1/4 & 1.001 & 0.052 & 0.050 & 931/1000 \\
+2/4 & 1.004 & 0.051 & 0.049 & 935/1000 \\
+3/4 & 1.001 & 0.052 & 0.050 & 937/1000 \\
log$k$ - binning & 1.002 & 0.051 & 0.050 & 927/1000 \\
$k_{\rm max}=0.30\,h{\rm Mpc}^{-1}$ & 1.002 & 0.051 & 0.051 & 934/1000 \\
\hline
QPM mocks:\\
~~~~~$\xi(s)$:\\
 fiducial & 1.001 & 0.051 & 0.052 & 361/400\\
 5$h^{-1}$Mpc & 1.000 & 0.050 & 0.051 & 355/400\\
 QPM cov  & 1.002 & 0.051 & 0.052 & 369/400\\
~~~~~$P(k)$:\\
fiducial  & 0.998 & 0.049 & 0.051 & 354/400 \\
QPM cov  & 0.999 & 0.049 & 0.049 & 359/400 \\
\hline
\label{tab:baomock}
\end{tabular}
\end{table}

For the correlation function, the mean $\alpha$ values are similar to those obtained when testing the mean of the mocks, as expected. We test bin sizes of 5$h^{-1}$Mpc and 8$h^{-1}$Mpc; the results are extremely similar and we thus choose 8$h^{-1}$Mpc binning as it represents the smaller data vector. The QPM and EZ mocks produce consistent results, suggesting the separate methods agree on the expected BAO signal-to-noise ratio in our DR14 quasar sample. In particular,  the results for the QPM mocks are insensitive to whether we use the QPM mocks or the EZ mocks to construct the covariance matrix used for the fits. This suggests our results are robust to uncertainties in how we construct the covariance matrix.

For 8$h^{-1}$Mpc correlation function binning, results using different bin centres are not perfectly correlated. Instead, the correlation factors, when shifting by each available 2$h^{-1}$Mpc shift, are close to 0.9 (the range is 0.88 to 0.92). Thus, slightly improved results are obtained when combining the results of these four bin centres; for Gaussian distributions, we would expect a 4 per cent improvement given four measurements with correlation 0.9. We combine the results by simply taking the mean of the four likelihoods, matching the approach of \cite{Ross15}. This combination increases the number of detections from 932 to 939 and reduces the standard deviation from 0.050 to 0.049. These results are labeled {`combined'} in Table \ref{tab:baomock} and represent our optimized method for measuring the correlation function BAO scale.

We find broadly consistent results in Fourier space. The fiducial power spectrum case recovers essentially the same number of detections (929 compared to 932) and has a slightly larger mean uncertainty and standard deviation. We test four bin centres for $P(k)$, shifting the centre by factors of 0.025$k$Mpc$^{-1}$. These results are more correlated than the $\xi(s)$ bin centre results, as they range 0.94 and 0.96. Combining across the bin centre results by taking the mean of the four likelihoods increases the number of detections to 941 while keeping the mean uncertainty and standard deviation unchanged. The power spectrum results are consistent whether the QPM or EZ mocks are being tested, whether a logarithmic or linear $k$-space binning is used, and whether the maximum $k$-vector for the analysis is set to $0.23$ or $0.30\,h{\rm Mpc}^{-1}$.

\begin{figure}
\includegraphics[width=84mm]{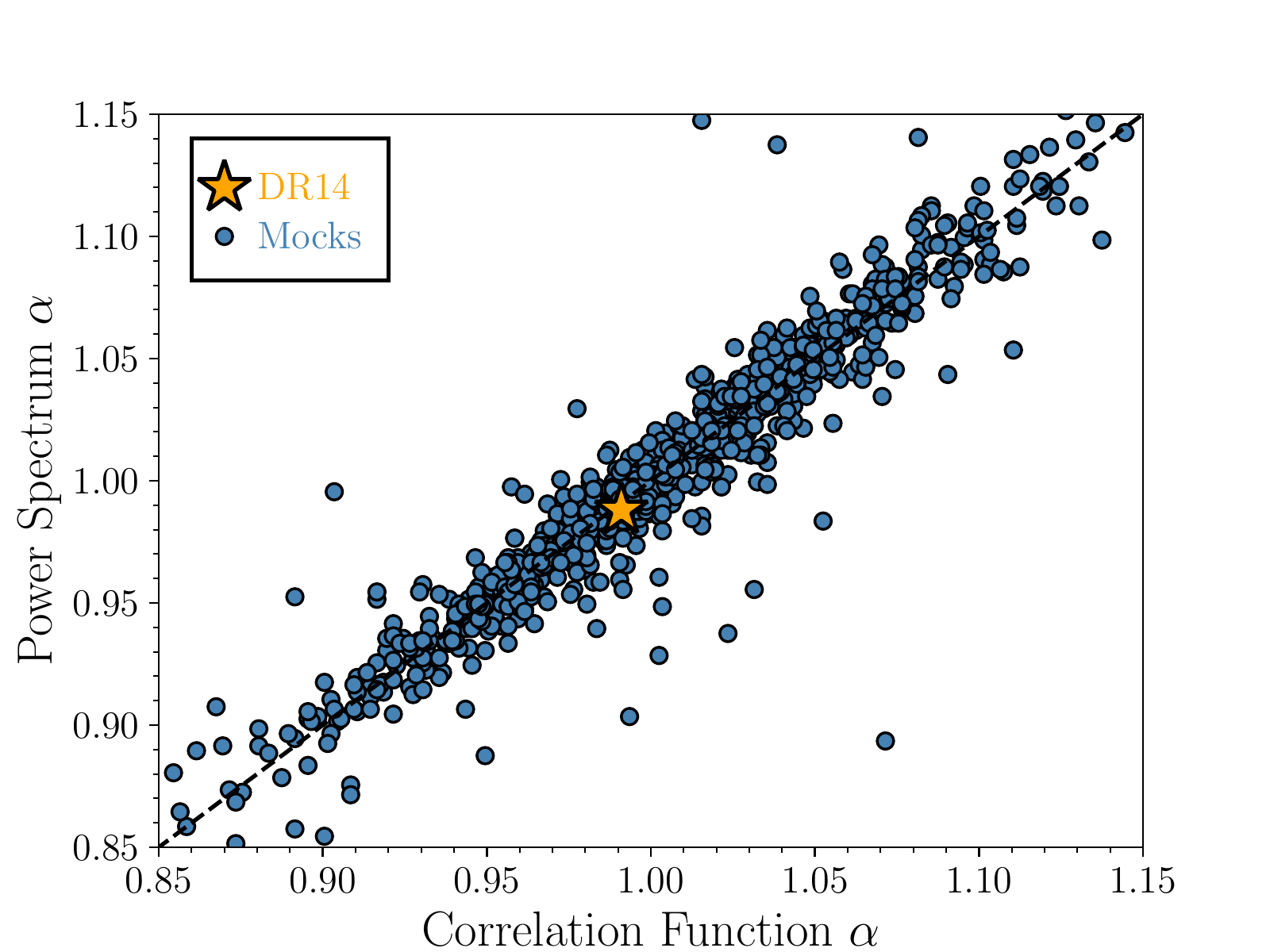}
\includegraphics[width=84mm]{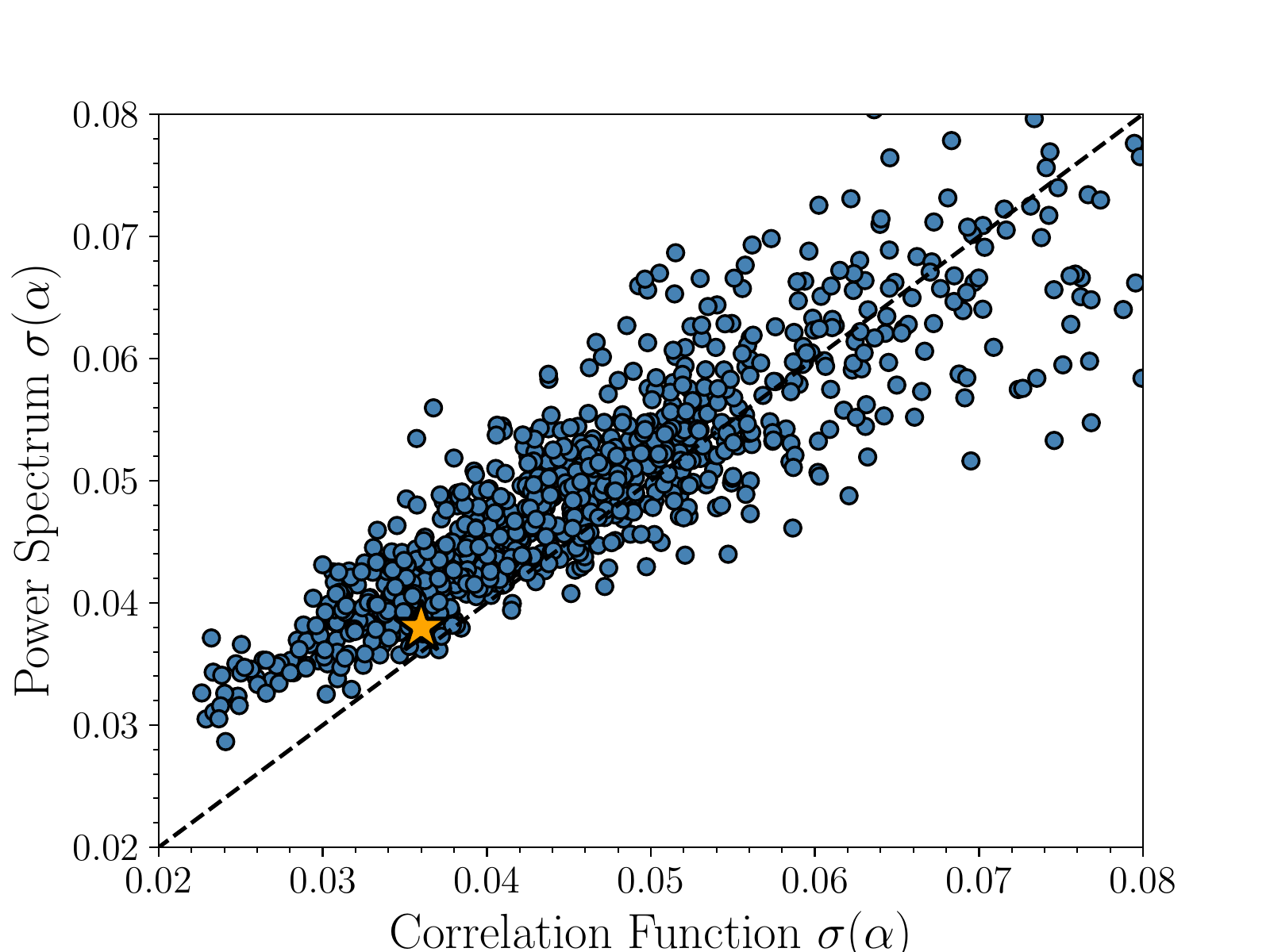}
  \caption{A comparison of power spectrum and correlation function BAO fit results. The results for the mocks are shown in steelblue circles, with the result for the DR14 quasar data indicated by an orange star. {\it Top panel}: The best-fit BAO parameter, $\alpha$. The two statistics reveal strongly correlated results, with a correlation coefficient 0.97. {\it Bottom panel}: The uncertainty on $\alpha$, recovered from the likelihood. The results are correlated, but the power spectrum uncertainties are drawn from a more narrow distribution. The differences between the power spectrum and correlation function results are clearly typical of our mock samples.
  }
  \label{fig:BAOpkxi}
\end{figure}

We compare the $P(k)$ and $\xi(s)$ results directly for the cases where we averaged across the bin centres for the EZmocks (denoted `combined' in Table \ref{tab:baomock}). The recovered $\alpha$ are plotted against each other in the top panel Fig. \ref{fig:BAOpkxi}, using steelblue circles. The results are strongly correlated (the correlation factor is 0.97) and are unbiased relative to each other. The bottom panel in Fig. \ref{fig:BAOpkxi} displays the uncertainty of the combined $P(k)$ and $\xi(s)$ results. The uncertainties are not as correlated as the best-fit $\alpha$, as the uncertainties obtained from the $P(k)$ likelihoods are more narrowly distributed than the $\xi(s)$ counterparts.
 
\begin{figure}
\includegraphics[width=84mm]{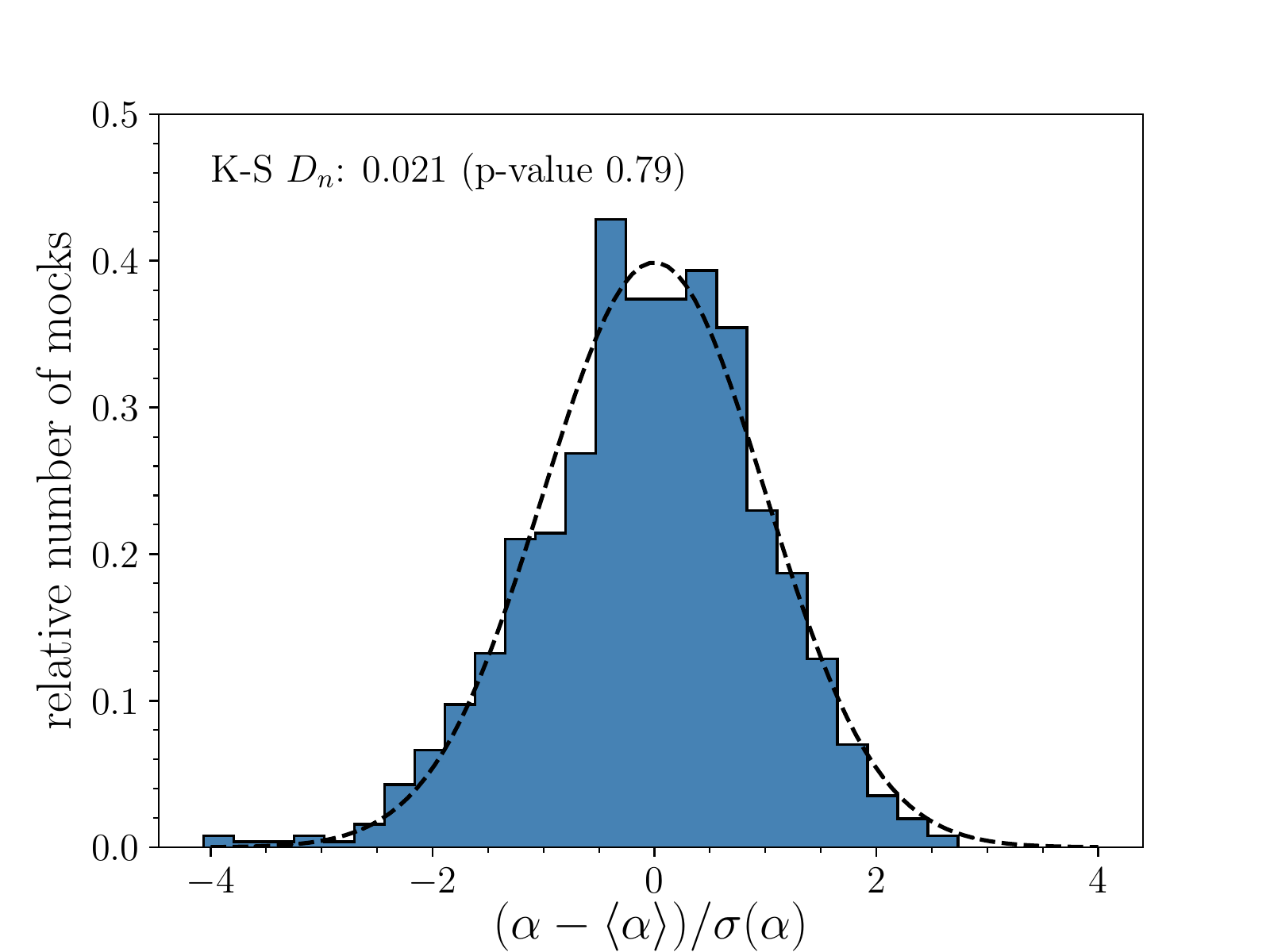}
  \caption{ The distribution of mock BAO results obtained from combining the power spectrum and correlation function BAO results, compared to a unit Gaussian. The low $D_n$ and the high $p$-value suggest that the uncertainties we obtain are as expected for a Gaussian distribution.
  }
  \label{fig:BAOks}
\end{figure}

Given that the combined $P(k)$ and $\xi(s)$ produce slightly different likelihoods, but with strong correlation in the maximum likelihood, we adopt the mean of the $P(k)$ and $\xi(s)$ results as a `consensus' measurement. Doing so, we obtain a slight increase in the number of detections (up to 944) and the standard deviation in the maximum likelihood value of $\alpha$ matches the mean uncertainty obtained from the likelihood. Further, $(\alpha-\langle \alpha\rangle)/\sigma(\alpha)$ is distributed similarly to a unit Gaussian distribution, as shown in Fig. \ref{fig:BAOks}. These results suggest that the uncertainty we obtain by combining the $P(k)$ and $\xi(s)$ likelihoods are indeed good estimates of the uncertainty on $\alpha$ for each mock realization. We copy this approach when obtaining our consensus results using the DR14 quasar data.

The tests presented in this Section define our procedure for obtaining BAO measurements from the data. The results suggest that there are no reasonable methodological choices that will affect our estimate of the uncertainty on our measurements by more than 10 per cent or bias our results by more than 0.1$\sigma$.

%section
\section{Results}
\label{sec:results}

%subsection
\subsection{Clustering Measurements}
\label{sec:clustering}
We measure the clustering of the DR14 quasar sample in the respective SGC and NGC regions, i.e., the window function is normalized in each respective region, and then the results are combined. In this section, we present the clustering measurements, in both configuration and Fourier space, comparing the results to each other and to the mock DR14 samples.

Fig. \ref{fig:xiNScompEZ} displays the spherically averaged redshift-space correlation function of the DR14 quasar sample for the data in the SGC (blue squares) and NGC (red diamonds). The dashed curves display the mean of the 1000 EZmock samples. The data in each region is broadly consistent with the mean of the mocks and with each other. Each recovers a good $\chi^2$ when tested against the mean of the mocks, both when the NGC and SGC are combined and compared individually. We use the sum of the respective covariance matrices to test the consistency between the NGC and SGC, over the same range of scales. The $\chi^2$ is 30.0 for these 24 measurement bins, suggesting the clustering of the two regions is consistent. 

The bottom panel of Fig. \ref{fig:xiNScompEZ} displays the results combining NGC and SGC (now with a solid black curve for the data) and includes the mean of the QPM mocks for an additional comparison. The $\chi^2$/dof, when comparing against both sets of mocks, is close to one and what appears to be the BAO feature can be seen at $s\sim100h^{-1}$Mpc. In configuration space, the DR14 quasar clustering is consistent with the expected signal and noise. The grey dashed line in the figure displays the clustering of the DR14 quasar sample when the systematic weights (defined in Section \ref{sec:weights}) for depth and Galactic extinction are not used. The $\chi^2$ in comparison to the mocks is labeled in parentheses in the figure; it is worse by $\Delta \chi^2 > 190$. This result is equivalent to having greater than a 13$\sigma$ effect, and can be compared to BOSS DR12, where the systematic weights had at most a 4$\sigma$ effect on the measured clustering \citep{Ross17}.

Fig. \ref{fig:p0compmock} displays the same information as in Fig. \ref{fig:xiNScompEZ}, but in Fourier space. As expected, there is a similar consistency between the NGC and the SGC results, compared to each other and to the mean of the mock samples. For this comparison, we have added a constant value to the mock results; otherwise the agreement would not be so excellent. The mean $P(k)$ of the mocks with no constant applied are shown by the dotted curves. We marginalize over a free constant term in the BAO analysis, and therefore believe this to be a fair comparison to evaluate the agreement, in terms of the signal relevant to the BAO measurement. Further, this difference in power clearly does not strongly affect the covariance matrix, as we obtained nearly identical results when fitting the QPM mocks when using either the QPM mocks or EZmocks to construct the covariance matrix. The effect of the systematic weights is negligible at scales $k>0.02h$Mpc$^{-1}$; the result without applying the weights is barely distinguishable (it is plotted with a grey, dashed curve) and the $\chi^2$ comparison to the mean of the mock samples changes by at most 5.2 (when compared to the QPM mocks). These results imply that, while the effect of the systematic weights has great total significance, any effect on the measurement of the BAO scale is likely to be negligible.

%subsection
\subsection{BAO Measurements}
\begin{figure}
\includegraphics[width=84mm]{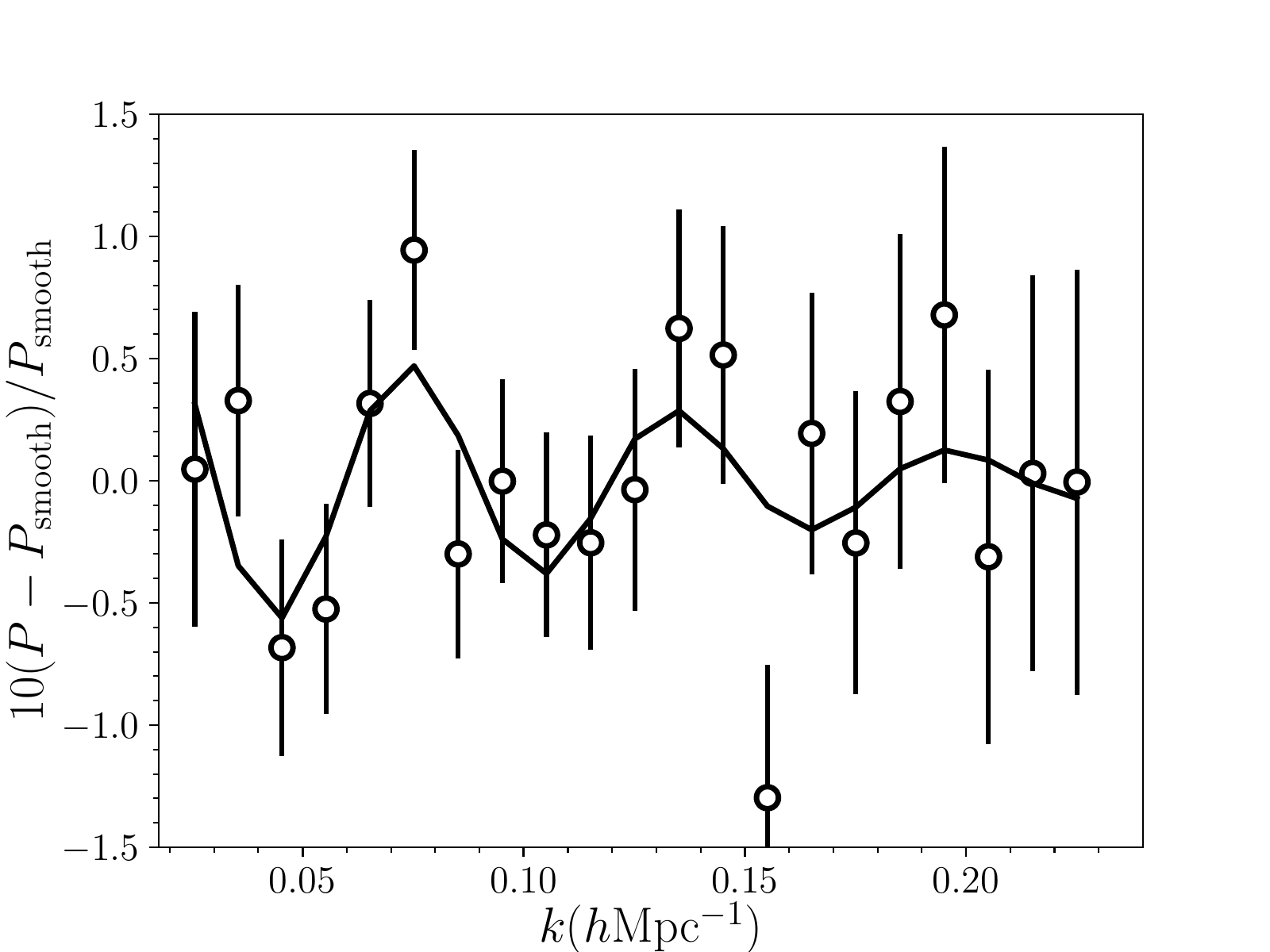}
\includegraphics[width=84mm]{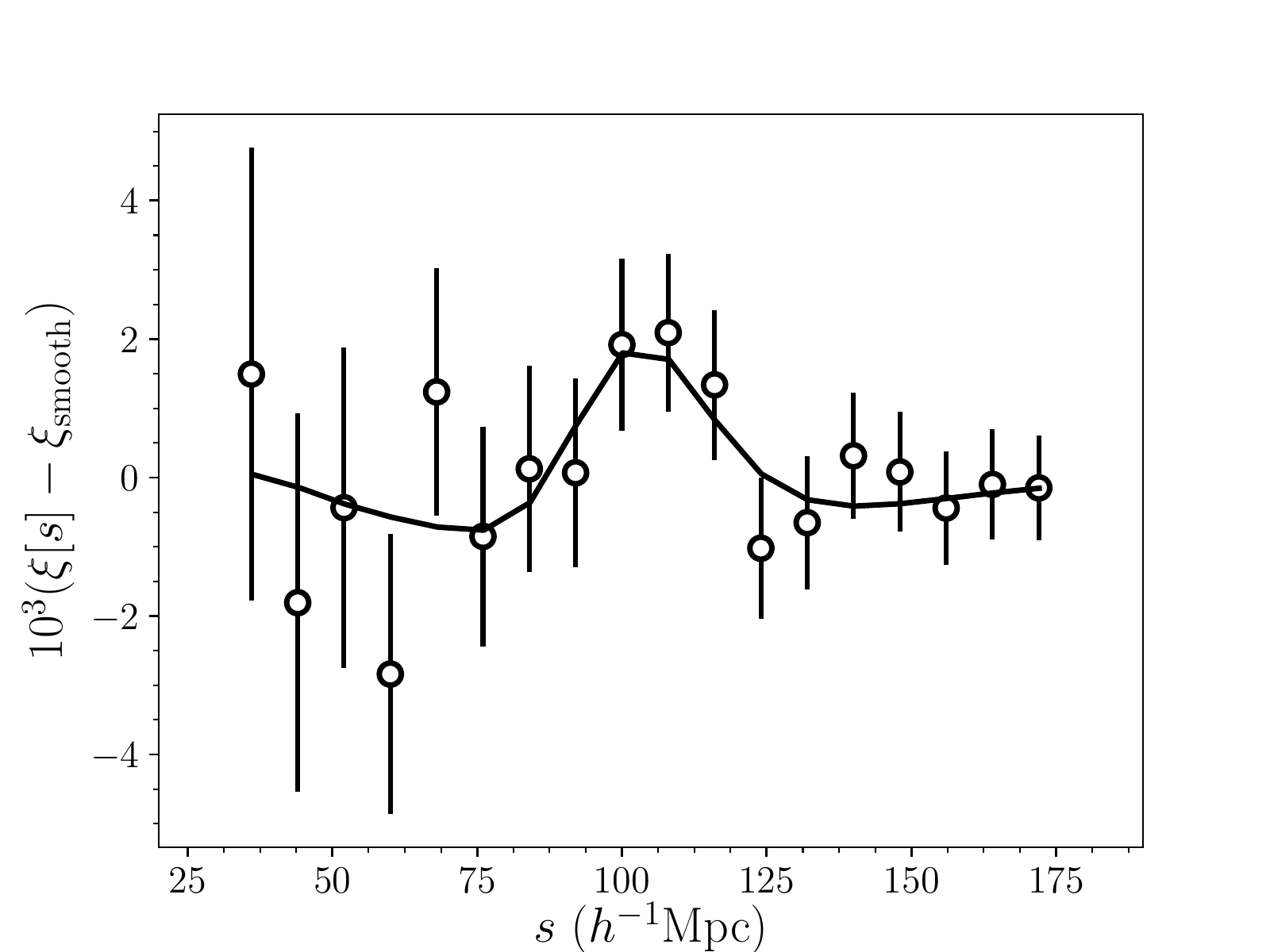}
 \caption{The eBOSS DR14 quasar spherically-averaged BAO signal, in Fourier- (top; $P(k)$) and configuration- (bottom; $\xi(s)$) space. In order to isolate the BAO feature, we have subtracted the smooth component of the best-fit model from the best-fit model and the measurements. In Fourier-space, we have additionally divided by the smooth component of the best-fit $P(k)$ model. Each clustering statistic prefers the BAO model to the smooth model at better than 2.5$\sigma$ and obtains a BAO distance measurement with a precision slightly greater than 4 per cent. }
 \label{fig:xiBAObf}
\end{figure}

Fig. \ref{fig:xiBAObf} displays the measured BAO feature in the eBOSS DR14 quasar sample, using our fiducial analysis choices. The top panel shows the Fourier-space result and the bottom panel the result in configuration space. The BAO feature has been isolated in each case by subtracting the smooth component of the best-fit model; for $P(k)$, we also divide the results by the smooth component. A clear BAO feature is visible in both spaces.

The statistics for the BAO measurement can be found in top rows of Table \ref{tab:baodata}. These are the `combined' results, where we have taken the mean likelihood across our four bin centres (as described in Section \ref{sec:mocktests}). For both $P(k)$ and $\xi(s)$, the $\chi^2$/dof is less than 1 and the precision is close to 4 per cent, with $\xi(s)$ obtaining somewhat better precision (3.7 compared to 4.0 per cent). The two measurements differ by  only 0.001 in $\alpha$. If anything, the agreement is surprisingly good. One can compare the orange star, representing our DR14 measurements, to the locus of mocks in the top panel of Fig. \ref{fig:BAOpkxi}. The bottom panel of the same figure displays the comparison of the uncertainties we recover from each measurement. Our results are more precise than the average results but are clearly within the locus of points thus suggesting they are consistent with any expectations provided by our tests on mocks. We combine the two likelihoods and obtain a precision of 3.8 per cent. Translating this result to a distance measurement yields $D_V(z=1.52)=3843\pm147 \left(r_{\rm d}/r_{\rm d, fid}\right)\ $Mpc.

\begin{figure}
\includegraphics[width=84mm]{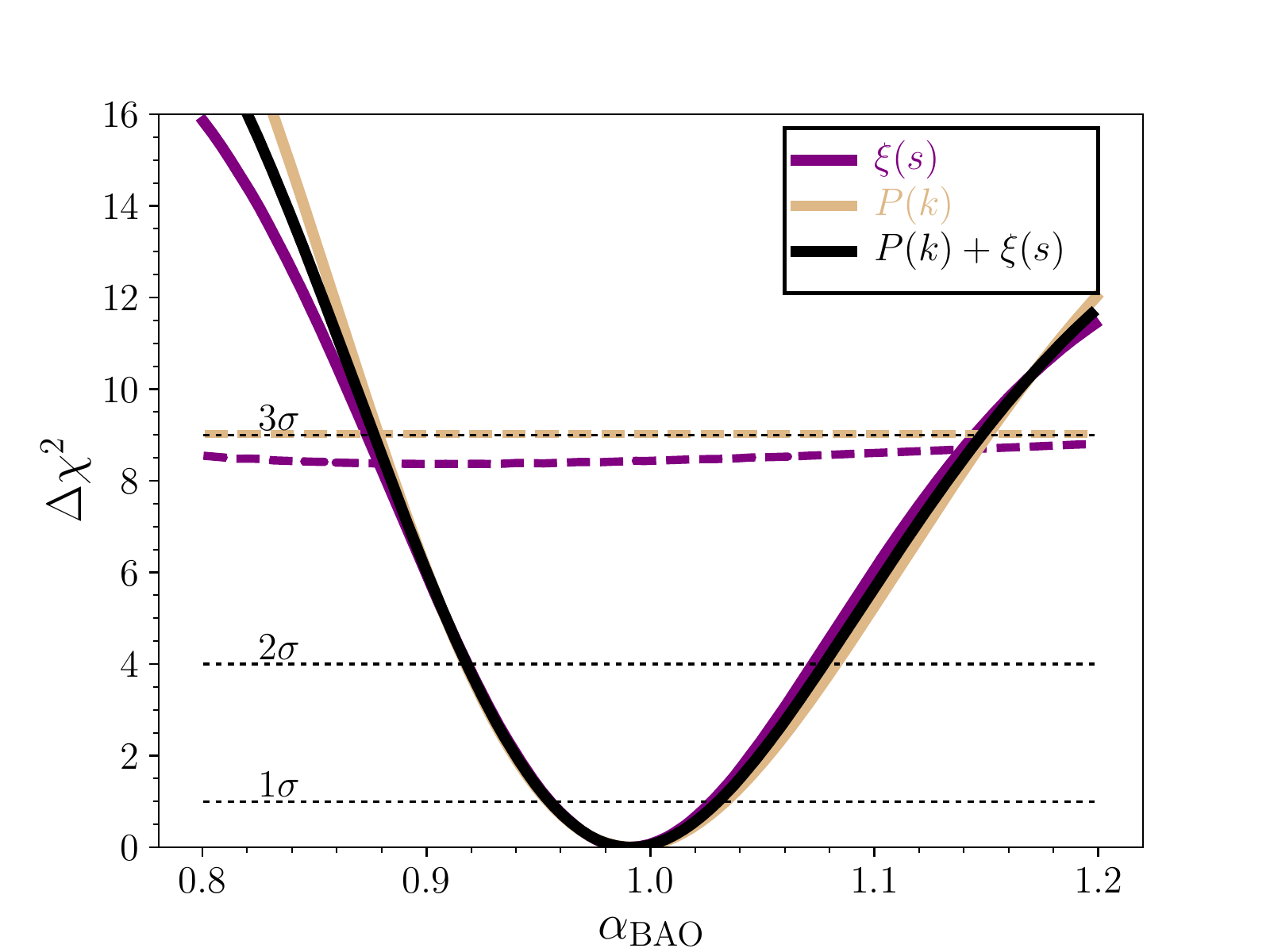}
  \caption{The solid curve displays the likelihood of the BAO parameter $\alpha$, in terms of $\Delta\chi^2$, recovered from the spherically averaged correlation function of the DR14 quasar sample. The dashed curve displays the same information for a no BAO model, where $\Delta\chi^2$ is determined by subtracting the minimum $\chi^2$ from the BAO model. The detection significance is slightly less than $3\sigma$ and the likelihood is slightly skewed compared to the Gaussian expectation. The purple curve displays correlation function results and the burlywood curve displays the power spectrum result. The black curve is the mean of the two likelihoods, which is what we use as our final measurement. 
  }
  \label{fig:xiBAOdet}
\end{figure}

Fig. \ref{fig:xiBAOdet} displays the likelihood and detection significance, in terms of $\Delta\chi^2$, derived from the spherically-averaged correlation function (purple), power spectrum (burlywood), and their mean (black). The dashed curve represents the no BAO model; one can observe that the detection significance is greater than 2.8$\sigma$ for both $P(k)$ and $\xi(s)$. All of the likelihoods are similarly skewed compared to a Gaussian, as large values of $\alpha$ are not rejected to the same extent as low values. The black curve represents the eBOSS DR14 quasar BAO distance measurement. For any cosmological tests, we recommend directly using this likelihood, which is publicly available\footnote{The BAO likelihood will be released publicly after the results are accepted by the journal for publication.}.

\begin{table}
\centering
\caption{Results for BAO fits to the DR14 quasar data. The fiducial $\xi$ case uses data with 8$h^{-1}$Mpc bin size and centres in the range $35 < s < 180h^{-1}$Mpc and the EZmock covariance matrix. For the $P$ the fiducial case uses data with linear binning of $0.01\,h{\rm Mpc}^{-1}$, in the range $0.02< k [h{\rm Mpc}^{-1}]<0.23$ and the EZmocks covariance matrix.  }
\begin{tabular}{lcc}
\hline
\hline
case & $\alpha$ & $\chi^2$/dof\\
\hline
{\bf DR14 Measurement} $P(k)+\xi(s)$ & $0.993\pm0.038$ & --\\
~~~~~$\xi(s)$ (combined) & $0.991\pm0.037$ & 6.2/13\\
~~~~~$P(k)$ (combined) & $0.992\pm0.040$ & 27.7/33\\
\hline
Robustness tests\\
~~~~$\xi(s)$:\\
fiducial & 0.996$\pm$0.039 & 8.6/13\\
+2 & 0.996$\pm$0.041 & 6.4/13\\
+4 & 0.984$\pm$0.033 & 3.2/13\\
+6 & 0.993$\pm$0.035 & 6.0/13\\
$Z_{\rm PCA}$ (combined) & 0.979$\pm$0.039 & 11.7/13\\
NGC & 0.975$\pm$0.054& 9.4/13\\
SGC & 1.014$\pm$0.057 & 18.9/13\\
QPM cov & 0.994$\pm$0.037 & 9.6/13\\
$\Delta s = 5h^{-1}$Mpc & 0.990$\pm$0.036 & 15.6/24\\
no $w_{sys}$ & 0.999$\pm$0.041 & 7.4/13\\
$50 < s < 150h^{-1}$Mpc & 0.997$\pm$0.042 & 7.9/8\\
$\Sigma_{\rm nl}=3.0h^{-1}$Mpc & 0.990$\pm$0.036 & 8.7/13\\
$\Sigma_{\rm nl}=9.0h^{-1}$Mpc & 1.004$\pm$0.045 & 9.6/13\\
$A_n = 0$ & 1.004$\pm$0.039 & 9.5/16\\
\smallskip
no $B$ prior & 0.997$\pm$0.037 & 8.8/13\\
~~~~$P(k)$:\\
$Z_{\rm PCA}$ (combined) & $0.980\pm 0.041$ & 28.2/33 \\
fiducial & $0.990\pm0.041$ & 30.1/33\\
$+1/4$ & $0.985\pm0.037$ & 25.4/33\\
$+2/4$ & $0.985\pm0.038$ & 25.0/33\\
$+3/4$ & $0.996\pm0.042$ & 30.3/33\\
NGC & $0.963\pm0.052$ & 15.8/16\\
SGC & $1.018\pm0.060$ & 13.8/16\\
QPM cov & $1.000\pm0.041$ & 29.7/33\\
log$k$ - binning & $0.997\pm0.042$ &  31.6/39\\
log$k$ - binning, $k_{\rm max}=0.30\,h{\rm Mpc}^{-1}$ & $1.002\pm0.040$ & 37.0/45 \\
no $w_{sys}$ & $0.992\pm0.045$ & 29.2/33\\
$k_{\rm max}=0.30\,h{\rm Mpc}^{-1}$ & $0.994\pm0.040$ & 53.3/47\\
$\Sigma_{\rm nl}=3\,h^{-1}{\rm Mpc}$ & $0.990\pm0.035$ & 29.6/33\\
$\Sigma_{\rm nl}=9\,h^{-1}{\rm Mpc}$ & $0.997\pm0.050$ & 30.2/33\\
$\Sigma_{\rm nl}= [6\pm3]\,h^{-1}{\rm Mpc}$ & $0.987\pm0.039$ & 29.9/32\\
$A_4\, A_5$ terms & $0.983\pm0.041$ & 20.6/29 \\ 
no-mask & $0.988\pm0.037$ & 28.4/33\\
\hline
\label{tab:baodata}
\end{tabular}
\end{table}

Robustness tests for our BAO measurements are shown in the bottom rows Table \ref{tab:baodata}. We find no particular causes for concern. Importantly, switching from our fiducial choice of redshift to $Z_{\rm PCA}$ shifts the recovered $\alpha$ by only 0.012 (less than one third $\sigma$) for both $\xi(s)$ and $P(k)$ and increases the mean of the $P(k)$ and $\xi(s)$ uncertainty from 0.038 to 0.040. These results have been combined across bin centres and can be compared directly to the DR14 results on the top lines of the table. The choice of redshift technique clearly does not have a strong impact on our results. \cite{Zarrouk17} will examine this issue more closely in the context of RSD measurements, but we note that, e.g., we find no clear differences in the linear bias or excess large-scale clustering obtained using either redshift.

Additional robustness tests should be compared to the `fiducial' results, which are the results for the fiducial choice of bin centre. The variations with bin centre for $\xi(s)$ (labeled $+x$) and $P(k)$ (labeled $+x/4$) are consistent with our findings testing our results with mocks; the mean of these four likelihoods is used as the DR14 $\xi(s)$ measurement, labeled `combined' in the top rows. Additional tests produce no changes that are greater than 0.2$\sigma$ for $\xi(s)$. For $P(k)$, the greatest change is {\bf only} 0.3$\sigma$ when switching to logarithmic binning and increasing $k_{\rm max}$ to 0.3$h$Mpc$^{-1}$. Finally, the results obtained from the independent NGC and SGC regions agree within 1$\sigma$, for both $\xi(s)$ and $P(k)$

The systematic weights have negligible effects on our measurements, despite their enormous (13$\sigma$) effect on the large-scale correlation function of the DR14 quasar sample. We simply recover less than a 0.1$\sigma$ shift in $\alpha$ and a 5 per cent increase in the uncertainty. The $\chi^2$ of the best-fit actually decreases (by 1), suggesting the effect of the depth and extinction systematics on our correlation function measurements are trivially accounted for with the polynomial terms in our BAO model.

For both $\xi(s)$ and $P(k)$, there is a slight correlation between the assumed $\Sigma_{\rm nl}$ and the recovered $\alpha$ that is less than $\sim$0.2$\sigma$ when testing in the range $\pm 3.0h^{1}$Mpc around our fiducial choice, with a similar impact on the size of the recovered uncertainty. For $P(k)$, we are able to marginalize over this parameter in the fit (using a Gaussian prior of $\pm 3.0h^{1}$Mpc) and we recover a result that matches the fiducial result with $\Sigma_{\rm nl}$ fixed at 6.0$h^{1}$Mpc to better than 0.1$\sigma$. Thus, not only does the choice of $\Sigma_{\rm nl}$ have a minor effect on our analysis, our choice for its fiducial value is sufficiently close to the best-fit value as to not make a difference in our results.

Overall, the robustness tests suggest that our results are insensitive to arbitrary choices in the analysis or the way the catalog was constructed. This is consistent with \cite{Ross17,2016arXiv161003506V} who showed that systematic uncertainties are small compared to the BOSS DR12 statistical uncertainties. The BOSS DR12 precision is a factor of four better than our own, and thus makes these systematic uncertainties negligibly small for our analysis. A separate systematic uncertainty is the possible shift in the acoustic peak due to a coupling of the quasar density field to the small relative velocity between baryons and cold dark matter at high redshift \citep{TH10,Dalal10,Yoo11,Slep15,Blaze16,Schmidt16}. This has been shown to be less than 0.5 per cent for low redshift galaxies \citep{Yoo13,Beutler16,Slepian16} and we expect it to be a minor effect for quasars with $z\sim1.5$, compared to our statistical uncertainty. Further study is warranted, especially as the statistical uncertainty will be considerably improved with future datasets.

\section{cosmological implications}
\label{sec:cosmo}

In this section, we briefly discuss the cosmological implications of our DR14 quasar BAO measurement of $D_V(z=1.52)=3843\pm147 \left(r_{\rm d}/r_{\rm d, fid}\right)\ $Mpc. We first present an updated BAO distance ladder and then demonstrate how this BAO distance ladder alone provides a powerful constraint on the geometry of the Universe.

\begin{figure}
\includegraphics[width=84mm]{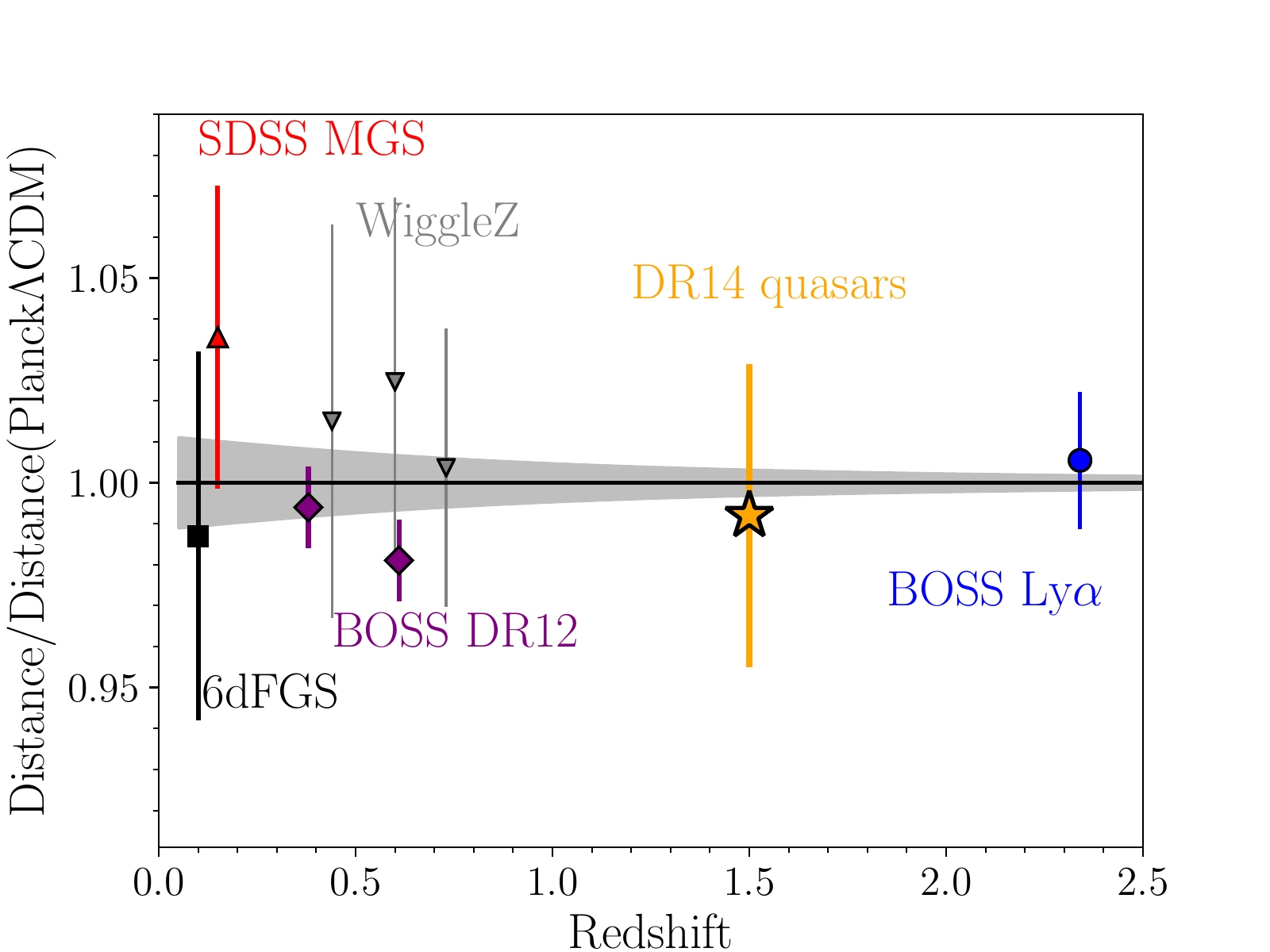}
  \caption{Spherically averaged BAO distance measurements ($D_V$) compared to the Planck $\Lambda$CDM prediction and extrapolated 68 per cent CL (grey region). The eBOSS DR14 quasar sample measurement is shown using a gold star. The additional measurements are described in the text.
  }
  \label{fig:BAOvPlanck}
\end{figure}

\begin{figure*}
\includegraphics[scale=0.4]{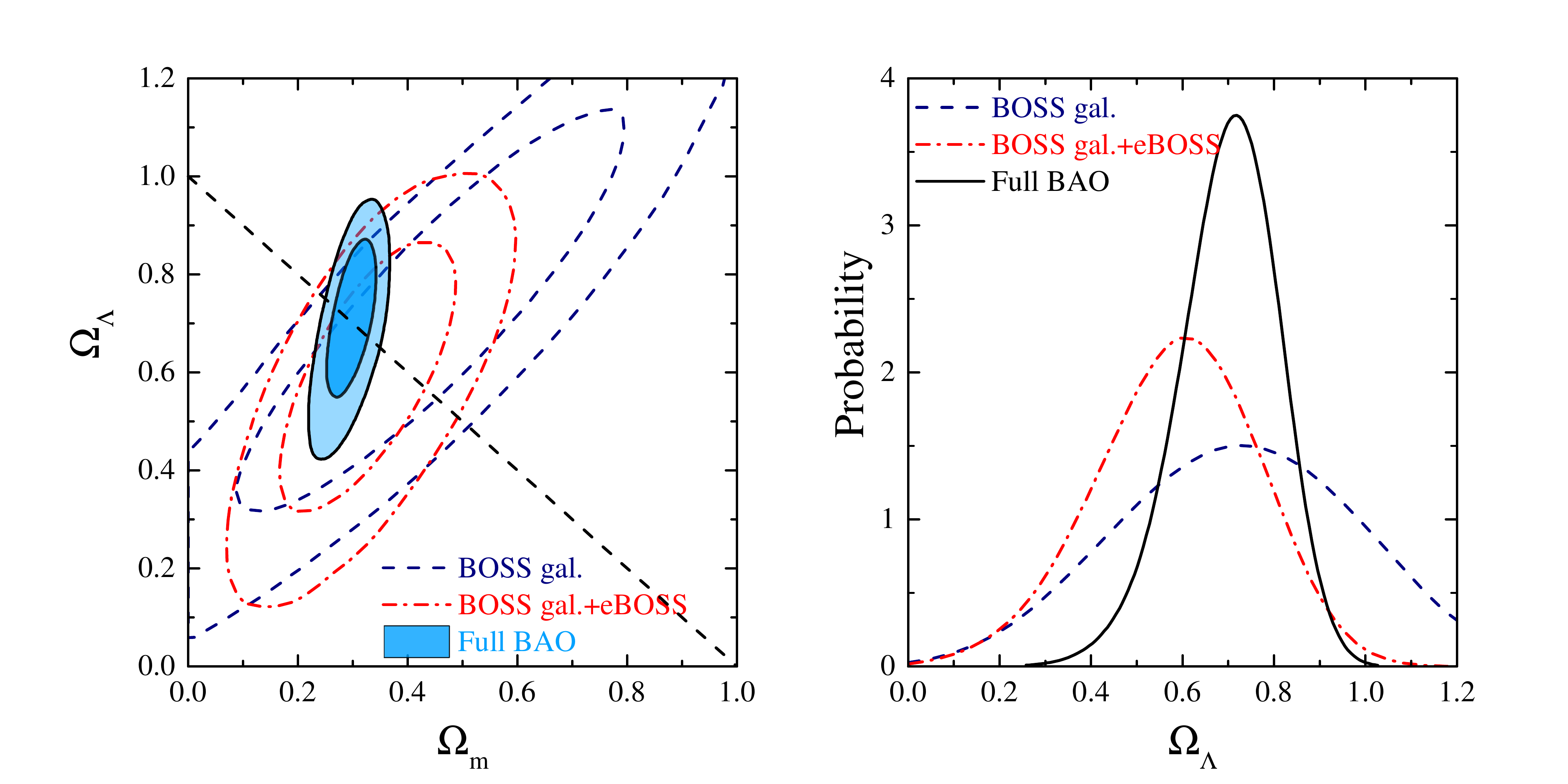}
 \caption{Left: The 68 and 95\% CL contour plots for $\Omega_{\rm m}$ and $\Omega_{\Lambda}$ using only the three sets of BAO data as illustrated in the legend. Here we assume only that the BAO scale is constant with redshift (and thus treat $r_{\rm d}$ as a nuisance parameter we marginalized over). The dashed line illustrates a flat Universe in which $\Omega_{\rm m} +\Omega_{\Lambda}=1$; Right: The one-dimensional probability distribution of $\Omega_{\Lambda}$ derived using three BAO datasets. Thus, the cosmology preferred by BAO distance scale measurements is flat $\Lambda$CDM and non-zero $\Lambda$ is preferred at 3.3$\sigma$ for the combination BOSS galaxies and our eBOSS DR14 quasar measurement and is preferred at 6.5$\sigma$ for the combination of all available BAO measurements (including BOSS Ly$\alpha$, which increases this preference the most). See text for further details. }
 \label{fig:cosmo}
\end{figure*}

Fig. \ref{fig:BAOvPlanck} displays our spherically averaged BAO measurements overplotted with the $\Lambda$CDM prediction from Planck \citep{Planck2015}, compared to various similar measurements in the literature: the 6dFGS result from \cite{6dF}, the SDSS MGS result from \cite{Ross15}, the BOSS DR12 results from \cite{Acacia}, the WiggleZ results from \cite{Kazin14}, and the BOSS Ly$\alpha$ from the combination of the \cite{Bautista17} DR12 Ly$\alpha$ auto-correlation and the \cite{Font14lya} measurement using the cross-correlation of the Ly$\alpha$ forest and quasars. Our measurement is in clear agreement with the expansion history predicted by Planck and other spherically averaged BAO distance measurements.

We have projected the combined Ly$\alpha$ results onto a $D_V$ measurement, which brings them into agreement with the Planck prediction. These Ly$\alpha$ measurements are in $>2\sigma$ tension when considering their sensitivity to $H(z)$ and $D_A(z)$ separately. We do not attempt to perform this decomposition with the DR14 quasar sample, as its signal-to-noise ratio makes it difficult to perform this decomposition robustly with BAO-only measurements. This will be done with future eBOSS quasar studies incorporating the RSD signal or using larger data sets.

We use the BAO distance ladder to constrain the geometry of the Universe. To do so, we assume only that the BAO feature has a constant co-moving size; we assume no knowledge of the physics that produced the feature. We use an open $\Lambda$CDM cosmology, which is parametrised using three parameters, \be P\equiv\{\Omega_{\rm m}, \Omega_{\Lambda}, H_0r_{\rm d}\} \ee where $\Omega_{\Lambda}$ denotes the fractional energy budget contributed by dark energy, and $H_0$ the Hubble parameter. This approach matches that recently presented in \cite{Aubourg15,Bautista17}. In this way, we are using only BAO measurements in order to test cosmology. We defer further study of the cosmological constraints afforded by our eBOSS DR14 quasar BAO measurement in combination with non-BAO data to future studies.

To obtain the constraint on $\Omega_{\rm m}$ and $\Omega_{\Lambda}$, which quantifies the cosmic geometry at the present epoch, we perform a Monte Carlo Markov Chain (MCMC) fitting using a modified version of {\tt CosmoMC} \citep{cosmomc02}, and marginalise over $H_0r_{\rm d}$. The datasets we use are as follows,

\begin{itemize}
\item BOSS galaxies: The anisotropic BAO measurement from BOSS DR12 presented in \cite{Acacia}; 
\item BOSS galaxies+eBOSS: The isotropic BAO measurement $D_V(z=1.52)=3843\pm147 \left(r_{\rm d}/r_{\rm d, fid}\right)\ $Mpc determined in this work combined with the BOSS DR12 BAO measurement;
\item Full BAO: BOSS galaxies+eBOSS combined with the anisotropic BAO measurement from the DR11 Lyman-$\alpha$ cross-correlation sample \citep{Font14lya} and the DR12 auto-correlation sample \cite{Bautista17}, and the isotropic BAO measurements using MGS \citep{Ross15} and 6dFGRS \citep{6dF} galaxy samples.
\end{itemize}

The 68 \& 95\% CL joint constraint on $\Omega_{\rm m}$ and $\Omega_{\Lambda}$, and the one-dimensional probability distribution of $\Omega_{\Lambda}$, are shown in Fig \ref{fig:cosmo}. The quasar BAO measurement in this work significantly improves the constraint, \ie,
\be \frac{\rm FoM_{\rm BOSS+eBOSS}}{\rm FoM_{\rm BOSS}} = 2.0 \ee where \be {\rm FoM}\propto 1/\sqrt{\rm det \ Cov(\Omega_{\rm m},\Omega_{\Lambda})} \ee denotes the Figure of Merit (FoM) of the geometric constraint of the Universe. The significance of $\Omega_{\Lambda}>0$, in other words, the existence of dark energy, is raised from $2.9\sigma$ to $3.4\sigma$ CL when the eBOSS quasar BAO is added to BOSS galaxies. Despite its relative lack in precision, the eBOSS DR14 quasar BAO measurement is able to provide a significant improvement over the BOSS galaxy BAO measurements alone as it provides a high-redshift constraint. Importantly, using all BAO measurements to date (the Full BAO) sample, we reach a $6.6\sigma$ detection of dark energy using BAO alone; this considerable improvement is mainly provided by the higher redshift and more precise BOSS Ly$\alpha$ measurements discussed earlier in this section\footnote{The DR14 measurement does not provide a significant improvement over what is achieved without it but with Ly$\alpha$; c.f. \cite{Bautista17}.}. Finally, all variations of the data set tested are in full agreement with a flat geometry.

BAO distance measurements continue to be in broad agreement with the flat $\Lambda$CDM model and the best-fit parameters from the Planck mission \citep{Planck2015} assuming this model.

%section, probably to be split in two
\section{Conclusions}
\label{sec:con}
We have used a sample of 147,000 quasars distributed over more than 2000 deg$^2$ in order to obtain the spherically-averaged BAO measurement $D_V(z=1.52)=3843\pm147 \left(r_{\rm d}/r_{\rm d, fid}\right)\ $Mpc. We have demonstrated this measurement is robust against a variety of methodological and observational concerns and choices, once again demonstrating BAO distance measurements to be one of the most robust observational probes of dark energy (as shown/discussed previously in, e.g., \citealt{Ross12,WeinbergDERev,Ross17,2016arXiv161003506V}).

These results demonstrate that the BAO signal in the distribution of quasars is consistent with expectations of basic LSS predictions. The clustering we measure and its BAO signal are consistent with that in our mock realizations. While the formation and evolution of quasars remains an active research field, they are clearly not so exotic as to greatly disturb the BAO signal.

We combine our result with previous, independent, BAO distance measurements to construct an updated BAO distance-ladder. Using these BAO data alone, we tested a $\Lambda$CDM model with free curvature, assuming only that the acoustic scale has a fixed comoving size. We found $\Omega_{\Lambda} > 0$ at 6.6$\sigma$ significance. Considering only BOSS galaxy and eBOSS quasar results, the significance remained greater than 3$\sigma$. All of our results are fully consistent with a flat geometry. BAO distance measurements, now across a broad range of redshifts, are in clear agreement with the flat $\Lambda$CDM cosmological paradigm.

This work represents the first cosmological analysis to be done with eBOSS quasar data. We expect numerous studies to follow, both with this catalog and with future, larger data sets. In particular, given the wide redshift coverage of the eBOSS quasar sample, there is potentially ample tomographic information along the radial direction. This aspect is crucial to reconstruct the history of the cosmic expansion and structure growth, which is key for the probe of dynamical dark energy \citep{Wang17BAO,Zhao17BAO,Zhao17DE}, modified gravity, and neutrino masses. We expect our DR14 results can be extended through the use of more optimal redshift-weighting methods for the BAO (e.g., \citealt{Zhu16,Wang17z}), redshift-space distortion (RSD) analyses, or their combination \citep{Ruggeri17}.  Additional BAO information can be extracted from higher-order statistics \citep{Slepian16bao}. Further, we anticipate the enormous volume probed by the entire eBOSS quasar sample will afford a precise measurement the signature of primordial non-Gaussianity. The final eBOSS quasar sample is expected to have approximately three times the volume of the DR14 sample, and will thus provide exciting improvements in the statistical precision of our BAO measurement, even without the expected methodological improvements.

The direct use of quasars as a tracer represents only one facet of the eBOSS program. Separate analyses of the eBOSS luminous red galaxy (LRG) and emission line galaxy (ELG) samples will measure BAO and RSD signal at redshift $z\sim$0.8, thereby filling the gap in redshift between BOSS galaxies and eBOSS quasars. Lyman$\alpha$ forest studies using eBOSS observations of quasars at $z>2.2$ will improve BAO measurements at $z\sim2.3$. Upcoming galaxy spectroscopic surveys will provide unprecedented precision; these include the Hobby-Eberly Telescope Dark Energy Experiment (HETDEX; \citealt{hetdex})\footnote{\url{http://hetdex.org/}}, Dark Energy Spectroscopic Instrument (DESI \citealt{DESI1,DESI2})\footnote{\url{http://desi.lbl.gov/}}, Prime Focus Spectrograph (PFS; \citealt{PFS})\footnote{\url{http://pfs.ipmu.jp/}} and the Euclid satellite mission \citep{Euclid}\footnote{\url{https://www.cosmos.esa.int/web/euclid/}}. These surveys will probe the Universe using multiple tracers including quasars, ELGs and LRGs. The work we have presented, and eBOSS studies in general, represent an exciting first step in obtaining a densely sampled BAO distance ladder to $z<3$.

%\clearpage
\section*{acknowledgements}
We thank the anonymous referee for comments that helped improve this paper.

AJR is grateful for support from the Ohio State University Center for Cosmology and ParticlePhysics.

HGM acknowledges support from the Labex ILP (reference ANR-10-LABX-63) part of the Idex SUPER, and received financial state aid managed by the Agence Nationale
de la Recherche, as part of the programme Investissements d'avenir under the reference ANR-11-IDEX-0004-02. 

GBZ is supported by NSFC Grant No. 11673025, and by a Royal Society Newton Advanced Fellowship.

RT acknowledges support from the Science and Technology Facilities Council via an Ernest Rutherford Fellowship (grant number ST/K004719/1)

CHC is grateful for support from Leibniz-Institut f\"{u}r Astrophysik Potsdam (AIP).

EB and PZ acknowledge support from the P2IO LabEx (ANR-10-LABX-0038).

JLT acknowledges support from National Science Foundation grant AST-1615997

YW is supported by the NSFC Grant No. 11403034

WJP acknowledges support from the UK Space Agency through grant ST/K00283X/1, and WJP acknowledges support from the European Research Council through grant {\it Darksurvey}, and the UK Science \& Technology Facilities Council through the consolidated grant ST/K0090X/1.

ADM was partially supported by the NSF through grant numbers 1515404 and 1616168.

IP acknowledges the support of the OCEVU Labex (ANR-11-LABX-0060) and the A*MIDEX project (ANR-11-IDEX-0001-02) funded by the ``Investissements d'Avenir" French government program managed by the AN

JPK acknowledges support from the ERC advanced grant LIDA.

GR acknowledges support from the National Research Foundation of Korea (NRF) through NRF-SGER 2014055950 funded by the Korean Ministry of Education, 
Science and Technology (MoEST), and from the faculty research fund of Sejong University in 2016.

Colours made possible by 
\url{http://matplotlib.org/examples/color/named_colors.html}; figures made colourblind-friendly (hopefully) by use of {\sc Color Oracle} software.

Funding for SDSS-III and SDSS-IV has been provided by
the Alfred P. Sloan Foundation and Participating Institutions.
Additional funding for SDSS-III comes from the
National Science Foundation and the U.S. Department of Energy Office of Science. Further information about
both projects is available at \url{www.sdss.org}.
SDSS is managed by the Astrophysical Research Consortium
for the Participating Institutions in both collaborations.
In SDSS-III these include the University of
Arizona, the Brazilian Participation Group, Brookhaven
National Laboratory, Carnegie Mellon University, University
of Florida, the French Participation Group, the German Participation Group, Harvard University,
the Instituto de Astrofisica de Canarias, the Michigan
State / Notre Dame / JINA Participation Group, Johns
Hopkins University, Lawrence Berkeley National Laboratory,
Max Planck Institute for Astrophysics, Max Planck
Institute for Extraterrestrial Physics, New Mexico State
University, New York University, Ohio State University,
Pennsylvania State University, University of Portsmouth,
Princeton University, the Spanish Participation Group,
University of Tokyo, University of Utah, Vanderbilt University,
University of Virginia, University of Washington,
and Yale University.

The Participating Institutions in SDSS-IV are
Carnegie Mellon University, Colorado University, Boulder,
Harvard-Smithsonian Center for Astrophysics Participation
Group, Johns Hopkins University, Kavli Institute
for the Physics and Mathematics of the Universe
Max-Planck-Institut fuer Astrophysik (MPA Garching),
Max-Planck-Institut fuer Extraterrestrische Physik
(MPE), Max-Planck-Institut fuer Astronomie (MPIA
Heidelberg), National Astronomical Observatories of
China, New Mexico State University, New York University,
The Ohio State University, Penn State University,
Shanghai Astronomical Observatory, United Kingdom
Participation Group, University of Portsmouth, University
of Utah, University of Wisconsin, and Yale University.

This work made use of the facilities and staff of the UK Sciama High Performance Computing cluster supported by the ICG, SEPNet and the University of Portsmouth.
This research used resources of the National Energy Research
Scientific Computing Center, a DOE Office of Science User Facility 
supported by the Office of Science of the U.S. Department of Energy 
under Contract No. DE-AC02-05CH11231.

\label{lastpage}

\end{document}